\documentclass[prb,twocolumn,tbtags,superscriptaddress,floatfix]{revtex4}
\usepackage{amsmath}
\usepackage{graphicx}
\usepackage{amssymb}
\usepackage{color,soul}
\usepackage[T1]{fontenc}
\usepackage{ae,aecompl}
\usepackage{natbib}

\begin{document}

\title{Qubit-based memcapacitors and meminductors}

\begin{abstract}
It is shown that superconducting charge and flux quantum bits (qubits) can
be classified as memory capacitive and inductive systems, respectively. We
demonstrate that such memcapacitive and meminductive devices offer
remarkable and rich response functionalities. In particular, when subjected
to periodic input, qubit-based memcapacitors and meminductors exhibit
unusual hysteresis curves. Our work not only extends the set of known
memcapacitive and meminductive systems to qubit-based devices but also
highlights their unique properties potentially useful for future
technological applications.
\end{abstract}

\author{Sergey N.~Shevchenko}
\affiliation{B.~Verkin Institute for Low Temperature Physics and Engineering, Kharkov
61103, Ukraine}
\affiliation{V.~Karazin Kharkov National University, Kharkov 61022, Ukraine}
\affiliation{CEMS, RIKEN, Saitama 351-0198, Japan}
\author{Yuriy V.~Pershin}
\affiliation{Department of Physics and Astronomy and Smart State Center for Experimental
Nanoscale Physics, University of South Carolina, Columbia, South Carolina
29208, USA}
\affiliation{CEMS, RIKEN, Saitama 351-0198, Japan}
\affiliation{Nikolaev Institute of Inorganic Chemistry SB RAS, Novosibirsk 630090, Russia}
\author{Franco~Nori}
\affiliation{CEMS, RIKEN, Saitama 351-0198, Japan}
\affiliation{Physics Department, University of Michigan, Ann Arbor, Michigan 48109-1040,
USA}
\maketitle

\section{Introduction}

There has recently been drastically increasing interest in electronic
circuit elements with memory, namely, memristive~[%
\onlinecite{chua71a,chua76a}], memcapacitive and meminductive~[%
\onlinecite{diventra09a}] systems (for a recent review see Ref.~[%
\onlinecite{pershin11a}]). In these resistive, capacitive and inductive
devices, the instantaneous response depends on the history of the signals
applied. While prominence has been given to memristive devices,
memcapacitive and (less frequently) meminductive devices are also
investigated. However, with few rare exceptions~[%
\onlinecite{Cohen12a,Gough15,Pfeiffer15a}], attention has been focused on
devices operating in the \textit{classical} regime. Therefore, it is
intriguing to find \textit{quantum} realizations, especially of
memcapacitive and meminductive systems, since these are not currently known.

In this regard, superconducting devices~[\onlinecite{Likharev}] are
attractive from several points of view. First of all, the past decade has
witnessed great progress in the area of superconducting qubits~[%
\onlinecite{Wendin07, You11, Xiang13}], which operate in the quantum regime.
Second, the relevance of \textit{classical} superconducting devices to the
area of memory circuit elements has already been established. Examples
include: $(i)$~phase-dependent conductance, interpreted as memristive
phenomenon~[\onlinecite{chua03a,Peotta14}]; $(ii)$~the voltage-history
dependence of the inductance (for more information on this meminductance,
see Appendix~A and Ref.~[\onlinecite{chua03a}]); $(iii)$~various hystereses
in different settings, such as, for example, the average voltage-current
hysteresis in the CRSJ model [\onlinecite{Likharev, Barone}], which can be
interpreted as a memristive phenomenon. In Appendix~A, the meminductance of
the Josephson junction is considered in detail to better explain some novel
aspects of the Josephson effect.

Surprisingly, it is not necessary to look very far to find examples of
\textit{quantum} superconducting memory devices. Indeed, the natural
candidate (a superconducting qubit~[\onlinecite{Wendin07, You11, Xiang13}])
is a quantum two-level system that, depending on the setting, offers a
memcapacitive or meminductive response. The goal of the present paper~%
\footnote{%
This paper extends preliminary work reported at the 2015 International
Symposium on Nanoscale Transport and Technology (ISNTT2015)~[%
\onlinecite{Pershin15}].} is to demonstrate the correspondence between
superconducting qubits and memory circuit elements. An interesting
distinctive feature of these quantum memory devices (compared to the
traditional ones such as considered in Ref.~[\onlinecite{martinez09a}]) is
their rich internal dynamics, stemming from the quantum internal dynamics of
qubits. In the past, many of such dynamical properties were demonstrated
experimentally, including coherent Rabi oscillations, Landau-Zener
tunneling, etc.~[\onlinecite{Wendin07,
Shevchenko10, You11, Xiang13}] We emphasize that while we consider
superconducting qubits, our approach can be extended to other types of
qubits.

Mathematically, memory circuit elements are defined by~[%
\onlinecite{diventra09a}]
\begin{eqnarray}
y(t) &=&g(\mathbf{x},u,t)u(t),  \label{y(t)} \\
\mathbf{\dot{x}} &=&\mathbf{f}(\mathbf{x},u,t).  \label{x_dot}
\end{eqnarray}%
Here, $u(t)$ and $y(t)$ are complementary constitutive circuit variables
denoting the input and the output of the system, $g$ is the generalized
response function, $\mathbf{x}$ is the set of variables describing the
internal state, and $\mathbf{f}$ is the vector function defining the
evolution of $\mathbf{x}$.

To be more specific, voltage-controlled memcapacitive systems~[%
\onlinecite{diventra09a}] are described by
\begin{eqnarray}
Q(t) &=&C_{\mathrm{M}}(\mathbf{x},V,t)V(t),  \label{CM} \\
\mathbf{\dot{x}} &=&\mathbf{f}(\mathbf{x},V,t),  \label{x_dot1}
\end{eqnarray}%
where the memcapacitance $C_{\mathrm{M}}$ is given by the relation between
the charge $Q$ and voltage $V$. Current-controlled meminductive systems~[%
\onlinecite{diventra09a}] are given by
\begin{eqnarray}
\phi (t) &=&L_{\mathrm{M}}(\mathbf{x},I,t)I(t),  \label{LM} \\
\mathbf{\dot{x}} &=&\mathbf{f}(\mathbf{x},I,t),  \label{x_dot2}
\end{eqnarray}%
where the meminductance $L_{\mathrm{M}}$ defines the relation between the
flux-linkage $\phi \equiv \int V\mathrm{d}t$ and the current $I$. We note
that Eqs.~(\ref{CM}-\ref{x_dot1}) and (\ref{LM}-\ref{x_dot2}) are particular
cases of Eqs.~(\ref{y(t)}-\ref{x_dot}).

In what follows, we show that the above equations match the equations for
certain expectation values calculated for qubit-based memcapacitors and
meminductors. In other words, the suggested devices behave \textit{on average%
} as classical memcapacitors and meminductors in some simple circuits
studied in this work. In what follows, such quantum-mechanically averaged
values are denoted with angular brackets. At the same time, the individual
measurements of the output of our devices will exhibit quantum uncertainty.
This uncertainty is a clear manifestation of the non-classical (quantum)
nature of our devices. This issue is also addressed in Ref.~[%
\onlinecite{Salmilehto16}], where a driven quantum system is described by
the respective average voltages and currents, of which the relation is
studied for the description of the quantum memristor operation.\emph{\ }%
Additionally, in more complex circuits than the ones considered here,
several qubit-based devices may form various non-trivial quantum states
(such as, e.g., entangled states) that would require a quantum approach to
describe the circuit dynamics. In any case, in this work the qubit-based
memcapacitors and meminductors are considered as quantum systems capable to
store quantum information.

This paper is organized as follows. In Sec.~\ref{sec2}, we present
descriptions of charge and flux qubits as memory circuit elements, showing
that their electrical response can be formulated in the form of Eqs.~(\ref%
{CM}) and (\ref{LM}), respectively. Then, in Sec.~\ref{sec3}, we discuss the
dynamics of the internal state variables of qubits. We show that the
equations of motion for the internal state variables can be written in the
form of Eqs.~(\ref{x_dot1},~\ref{x_dot2}). Simulation results are described
in Sec.~\ref{sec4}, which presents various types of hysteretic loops.
Finally, we conclude in Sec.~\ref{sec5}. Details of calculations are
presented in the Appendices. Moreover, the quantum uncertainty of
measurements is discussed in the last Appendix.

\section{Qubits as memory devices}

\label{sec2}

\begin{figure}[t]
\includegraphics[width=8cm]{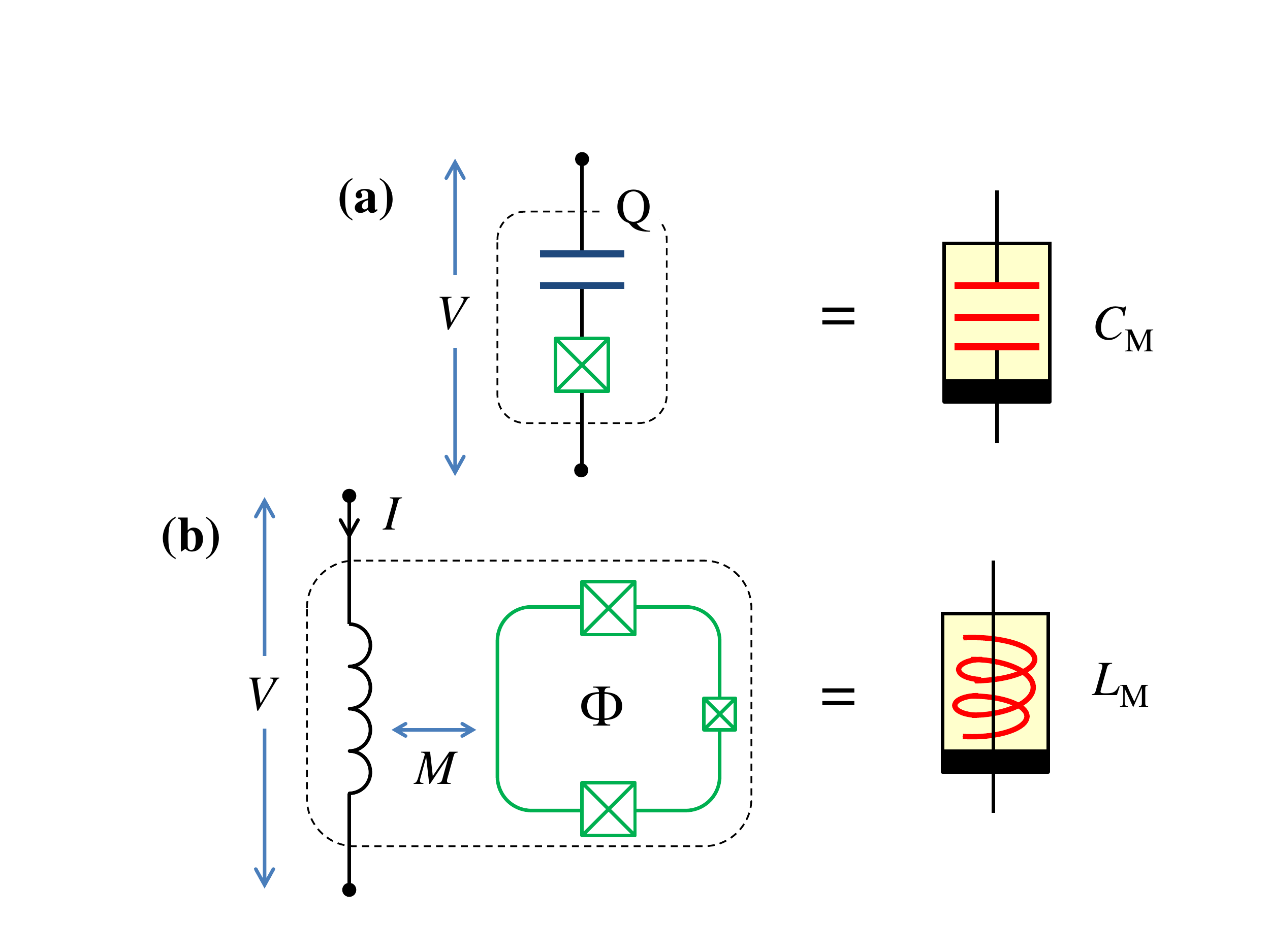}
\caption{Superconducting qubits as memory circuit elements.
(a) The charge qubit implements a memcapacitive system. (b) A coil
inductively coupled to a flux qubit forms an effective meminductive system.
The crossed boxes denote Josephson junctions, the circuit symbols of
memcapacitor (top) and meminductor (bottom) are shown on the right.}
\label{Fig:schemes}
\end{figure}

We will focus on charge and flux qubits, showing that they belong to the
general classes of memcapacitive and meminductive systems. For this purpose,
we cast the qubit equations in the form of Eqs.~(\ref{CM}-\ref{x_dot1}) and (%
\ref{LM}-\ref{x_dot2}), respectively, thus identifying the internal state
variables, response and evolution functions. It is interesting that the
equations for charge and flux qubits can be written exactly in the same
form, when we treat these structures as memory circuit elements. The circuit
elements and notations are summarized in Fig.~\ref{Fig:schemes} and Table~%
\ref{t}.

Below, we use a semiclassical approach, where the quantum-mechanical
evolution of the qubit is considered in the presence of the classical input $%
u(t)$. The system output $y(t)$ is calculated as an expectation value. Such
model assumes the input and output to be described by coherent states,
involving many photons. For more details on the semiclassical approximation,
see, e.g., Ref.~[\onlinecite{Shevchenko14}].

In what follows, we consider both the case when the dissipative environment
can be disregarded (good isolation; the system can be described by the
Liouville equation), and also the case with significant dissipation (which
is introduced phenomenologically and may include the effect of the
measurement apparatus; in this case, the system is described using the
dissipative Bloch equation). The Bloch equation includes the effect of
non-zero temperature $T$ as well as the relaxation and decoherence rates, $%
\Gamma _{1}$ and $\Gamma _{2}$. We note that neglecting these rates, $\Gamma
_{1}=\Gamma _{2}=0$, reduces the Bloch equation to the Liouville equation.

\begin{table}[b]
\caption{Charge and flux qubits as memcapacitive and meminductive systems.}
\label{t}\renewcommand{\arraystretch}{1.9}
\begin{ruledtabular}
\begin{tabular}{ccc}
    $ $& Charge qubit & Flux qubit \\
    \hline  $u(t)$ & $V(t)$ & $I(t)$ \\
    \hline  $y(t)$ & $Q$ & $\phi =\int V\mathrm{d}t$ \\
    \hline  $g$ & $C_{\mathrm{M}}=C_{\mathrm{geom}}-\frac{eC_{\mathrm{g}}}{VC_{\Sigma }}\left\langle \sigma
_{z}\right\rangle$ & $L_{\mathrm{M}}=L-\frac{MI_\mathrm{p}}{I}\left\langle \sigma
_{z}\right\rangle$ \\
    \hline  $\mathbf{x}$ & $\left(X,Y,Z\right) ^{\top }$ & $\left(X,Y,Z\right) ^{\top }$ \\
    \hline $f$ & $ \mathbf{B\times x}-\Gamma \mathbf{(x}-\mathbf{x}_{0})$ & $ \mathbf{B\times x}-\Gamma\mathbf{(x}-\mathbf{x}_{0})$ \\
    \hline  Type & Memcapacitive system &Meminductive system \\
    \end{tabular}
\end{ruledtabular}
\end{table}
\renewcommand{\arraystretch}{1}

\subsection{Charge qubit}

Consider first the superconducting charge qubit, Fig.~\ref{Fig:schemes}(a).
Its main part, the so-called Copper-pair box, is formed by a gate capacitor $%
C_{\mathrm{g}}$ and a Josephson junction with a capacitance $C_{\mathrm{J}}$%
. The superconducting island between these two capacitors has the total
capacitance $C_{\Sigma }=$ $C_{\mathrm{g}}+C_{\mathrm{J}}$, and is
characterized by the average charge $-2e\left\langle n\right\rangle $, where
$-e$ is the electron charge and $\langle n\rangle $ is the number of Copper
pairs on the island. The island is assumed to be biased by a voltage, which,
in general, contains both time-dependent, $V(t)$, and dc, $V_{\mathrm{dc}}$,
components. From electrostatic considerations, one finds the charge on the
external plate of the gate capacitor
\begin{equation}
Q(t)=\frac{C_{\mathrm{g}}C_{\mathrm{J}}}{C_{\Sigma }}\left[ V(t)+V_{\mathrm{%
dc}}\right] -\frac{C_{\mathrm{g}}}{C_{\Sigma }}2e\left\langle n\right\rangle
.  \label{Q}
\end{equation}

%One can introduce then the effective differential capacitance\onlinecite%
%{Sillanpaa05, Duty05, Johansson06} $C_{\mathrm{eff}}=dQ/dV$ and this would
%give that this consists of the geometric component $C_{\mathrm{geom}}=C_{%
%\mathrm{g}}C_{\mathrm{J}}/C_{\Sigma }$ and the so-called quantum capacitance
%\ $C_{\mathrm{Q}}=-\frac{C_{\mathrm{g}}}{C_{\Sigma }}2e\frac{\partial
%\left\langle n\right\rangle }{\partial V}$.
%For our purposes it is rather more convenient to consider the capacitance
%according to its standard definition, Eq.~(\ref{CM}).

In the two-level approximation~[\onlinecite{Sillanpaa05, Duty05, Johansson06}%
], the charge qubit Hamiltonian is written as
\begin{equation}
H=-\frac{\Delta }{2}\sigma _{x}-\frac{\varepsilon }{2}\sigma _{z},\text{ \ \
}\varepsilon =\varepsilon _{0}+\varepsilon _{1}(t),  \label{H}
\end{equation}%
where the energy bias $\varepsilon =E_{\mathrm{C}}(2n_{\mathrm{g}}-1)$ is
defined by the island charging energy $E_{\mathrm{C}}=(2e)^{2}/2C_{\Sigma }$
and the dimensionless gate voltage $n_{\mathrm{g}}=C_{\mathrm{g}}V/2e$. The
tunneling amplitude $\Delta =E_{\mathrm{J}}$ is given by the Josephson
energy of the contact, and the $\sigma _{i}$ stand for the Pauli matrices in
the charge representation. Using $n=(1+\sigma _{z})/2$ for the charge
operator, Eq.~(\ref{Q}) can be written in the form \footnote{%
We emphasize that there is a difference between the memcapacitance, defined
in Eq.~(\ref{QQ}), and the effective differential capacitance $C_{\mathrm{eff%
}}=\partial Q/\partial V$ often used in the literature~[%
\onlinecite{Sillanpaa05,
Duty05, Johansson06}]. The latter consists of the geometric component $C_{%
\mathrm{geom}}$ and the so-called quantum capacitance $C_{\mathrm{Q}%
}=-\left( C_{\mathrm{g}}/C_{\Sigma }\right) 2e\partial \left\langle
n\right\rangle /\partial V$.}
\begin{equation}
Q=C_{\mathrm{geom}}V-\frac{eC_{\mathrm{g}}}{C_{\Sigma }}\left\langle \sigma
_{z}\right\rangle \equiv C_{\mathrm{M}}(\mathbf{x},V)V,  \label{QQ}
\end{equation}%
where $C_{\mathrm{geom}}=C_{\mathrm{g}}C_{\mathrm{J}}/C_{\Sigma }$, and $%
\mathbf{x}$ stands for the set of parameters describing the time evolution
of the charge in the Cooper-pair box [through the $\left\langle \sigma
_{z}\right\rangle $ term]. In Eq.~(\ref{QQ}), for the sake of clarity, we
have eliminated the time-independent terms, by choosing the dc bias such
that $C_{\mathrm{J}}V_{\mathrm{dc}}/e=1$.
%In practice, it can be compensated by adding constant dc bias.
Clearly, Eq.~(\ref{QQ}) is equivalent with Eq.~(\ref{CM}). In this way, the
charge qubit can be considered as a \textit{memcapacitive} system.

Assuming a periodic input signal of amplitude $V_{A}$, Eq.~(\ref{QQ}) can be
presented in the dimensionless form as
\begin{equation}
\frac{Q(t)}{C_{\mathrm{geom}}V_{\mathrm{A}}}=\frac{V(t)}{V_{\mathrm{A}}}-%
\frac{e}{C_{\mathrm{J}}V_{\mathrm{A}}}\left\langle \sigma _{z}\right\rangle .
\label{memcapacitor}
\end{equation}%
In particular, taking $V(t)=V_{\mathrm{A}}\mathnormal{\sin }\omega t$, one
can find $\varepsilon _{1}(t)=A\mathnormal{\sin }\omega t$, where $A=2eC_{%
\mathrm{g}}V_{\mathrm{A}}/C_{\Sigma }$.

\subsection{Flux qubit}

Next, we consider the flux qubit coupled via the mutual inductance $M$ to
the inductor $L$ biased by the current $I$, Fig.~\ref{Fig:schemes}(b). In
this arrangement, the electrical response of the coil depends on the qubit
state. As the qubit state has a memory on the history of signals applied (to
the coil), it is natural to describe the entire system as an inductor with
memory, namely, a \textit{meminductive} system~[\onlinecite{diventra09a}]
operating in the quantum regime. Previously, it was demonstrated~[%
\onlinecite{Cohen12a}] that an $RCL$ contour inductively coupled to an
inductor represents a classical meminductive system.

The flux qubit is a superconducting ring with three Josephson junctions~[%
\onlinecite{Makhlin02, Wendin07}]. The two qubit states correspond to
persistent currents in the ring in the clockwise and counterclockwise
directions. The persistent current amplitude is $I_{\mathrm{p}}$. The ring
is pierced by a magnetic flux $\Phi $ with both ac, $\Phi _{\mathrm{ac}}$,
and dc, $\Phi _{\mathrm{dc}}$, components. The former is introduced by the
ac current in the inductor $L$, and the latter can be created by the dc
current in the same or in a separate inductor. In the two-level
approximation~[\onlinecite{Makhlin02, Wendin07}], the flux qubit is also
described by the Hamiltonian~(\ref{H}), where now the parameters have the
following meaning: $\Delta $ is the tunneling amplitude, $\sigma _{i}$ are
the Pauli matrices in the flux representation, $\varepsilon _{0(1)}=2I_{%
\mathrm{p}}\Phi _{0}f_{\mathrm{dc(ac)}}$ is the constant (time-dependent)
part of the bias defined by the dc (ac) component of the magnetic flux
through the qubit loop, $\Phi _{0}=h/(2e)$ is the magnetic flux quantum, $f_{%
\mathrm{dc}}=\Phi _{\mathrm{dc}}/\Phi _{0}-1/2$, and $f_{\mathrm{ac}%
}=MI(t)/\Phi _{0}$.

The electromotive force in the coil is given by $\mathcal{E}=-\dot{\Phi}_{%
\mathrm{c}}-\dot{\Phi}_{\mathrm{q}}$, where $\Phi _{\mathrm{c}}$ and $\Phi _{%
\mathrm{q}}$ are the magnetic fluxes through the inductor $L$ due to the
current $I$ in the coil and due to the qubit's current $I_{\mathrm{q}}$,
respectively. This can be rewritten for the voltage across the coil
\begin{equation}
V=L\dot{I}+M\dot{I_{\mathrm{q}}}.  \label{V}
\end{equation}%
Integrating Eq.~(\ref{V}) over time and using $I_{\mathrm{q}}=-I_{\mathrm{p}%
}\left\langle \sigma _{z}\right\rangle $ one obtains the expression for the
flux-linkage $\phi $ in the form of Eq.~(\ref{LM})
\begin{equation}
\phi =LI-MI_{\mathrm{p}}\left\langle \sigma _{z}\right\rangle \equiv L_{%
\mathrm{M}}(\mathbf{x},I)I.  \label{fi2}
\end{equation}%
We note that Eq.~(\ref{fi2}) also nominally coincides with Eq.~(\ref{QQ})
(see also the generalized notations in the Table~\ref{t}). In Eq.~(\ref{fi2}%
), the vector $\mathbf{x}$ stands for a set of parameters defining the qubit
state through $\left\langle \sigma _{z}\right\rangle $.

Finally, let us assume that the ac component of the current is $I_{A}\sin
\omega t$. Then, $f_{\mathrm{ac}}(t)=MI_{A}\sin \omega t/\Phi _{0}$, so that
${\varepsilon _{1}}(t)=A\sin \omega t$ with $A=2MI_{\mathrm{p}}I_{\mathrm{A}%
} $. In the dimensionless form, Eq.~(\ref{fi2}) can be written as
\begin{equation}
\frac{\phi (t)}{LI_{\mathrm{A}}}=\frac{I(t)}{I_{\mathrm{A}}}-\frac{MI_{%
\mathrm{p}}}{LI_{\mathrm{A}}}\left\langle \sigma _{z}\right\rangle .
\label{meminductor}
\end{equation}

\section{Dynamics of the Internal State Variables}

\label{sec3}

In the previous section we obtained relations for the memcapacitance and
meminductance, Eqs.~(\ref{QQ}) and (\ref{fi2}), in the form of Eq.~(\ref%
{y(t)}) with $g(\mathbf{x},u,t)$ defined by $\left\langle \sigma
_{z}\right\rangle $. The dimensionless forms of these expressions, Eqs.~(\ref%
{memcapacitor}) and (\ref{meminductor}), can be written in a unified form
\begin{equation}
\frac{y(t)}{y_{0}}=\frac{u(t)}{u_{0}}-\varkappa \left\langle \sigma
_{z}\right\rangle ,  \label{u(t)}
\end{equation}%
where, comparing with Eqs.~(\ref{memcapacitor}) and (\ref{meminductor}), one
can easlily identify $y_{0}$, $u_{0}$, and $\varkappa $ for the respective
two cases. Let us now clarify what are the variables that form the vector $%
\mathbf{x}$\ and define $\left\langle \sigma _{z}\right\rangle $.

Previously, the Hamiltonians and Pauli matrices $\sigma _{i}$ were defined
in the physical bases, which are the charge basis for the charge qubit and
the current basis for the flux qubit. In these bases, $\left\langle \sigma
_{z}\right\rangle $ provides the difference between the probabilities of the
two charge states and of the two current directions for the charge and flux
qubits, respectively. In order to describe the quantum dynamics of a qubit,
one has to take into account the dissipative processes. This can be done in
the framework of the Bloch equation~[\onlinecite{Blum, Makhlin02, Wendin07}%
]. Since the Bloch equation defines the relaxation in the energy
representation, one has to change to this basis (see Appendix B for more
details).

Let the qubit density matrix in the energy representation be parameterized
as $\rho =\frac{1}{2}(1+\mathbf{x\sigma })$ with $\mathbf{x=}\left(
X,Y,Z\right) ^{\top }$ being the so-called Bloch vector. The Bloch vector
thus plays the role of the \textit{internal state variables} of qubits.
Changing from the physical representation to the energy one, we obtain%
\begin{equation}
\left\langle \sigma _{z}\right\rangle =-\frac{\Delta }{\Delta E}X+\frac{%
\varepsilon _{0}}{\Delta E}Z,  \label{sigmaz}
\end{equation}%
with $\Delta E\equiv \hbar \omega _{\mathrm{q}}=\sqrt{\Delta
^{2}+\varepsilon _{0}^{2}}$. Equation~(\ref{sigmaz}) describes how the
response function $g$ in Eqs.~(\ref{QQ}) and (\ref{fi2}) depends on the
components of the Bloch vector $\mathbf{x}$. Moreover, for the sake of
simplicity, we assume that the two phenomenological relaxation rates
entering the Bloch equation are the same (see Appendix B), namely, $\Gamma
_{2}=\Gamma _{1}\equiv \Gamma $. In this way, one can write the Bloch
equation as
\begin{equation}
\mathbf{\dot{x}}=\mathbf{f}(\mathbf{x},u)\equiv \mathbf{B\times x}-\Gamma
\mathbf{(x}-\mathbf{x}_{0}),  \label{eq4x}
\end{equation}%
where%
\begin{eqnarray}
\mathbf{B} &=&\left( B_{x},0,B_{z}\right) ^{\top },\quad \mathbf{x}%
_{0}=\left( 0,0,Z_{0}\right) ^{\top }, \\
B_{x} &=&\frac{\Delta }{\Delta E}\frac{\varepsilon _{1}(t)}{\hbar }\equiv
2\Omega _{\mathrm{R}}^{(0)}\sin \omega t, \\
B_{z} &=&-\omega _{\mathrm{q}}-\frac{\varepsilon _{0}}{\Delta }B_{x},\quad
\Omega _{\mathrm{R}}^{(0)}=\frac{\Delta A}{2\hbar \Delta E},  \label{OR}
\end{eqnarray}%
and $Z_{0}=\tanh (\Delta E/2k_{\mathrm{B}}T)$ describes the equilibrium
energy level populations.

Equation~(\ref{eq4x}) corresponds to the generic equation~(\ref{x_dot}) and,
together with Eqs.~(\ref{QQ}) and (\ref{fi2}) [which are in the form of Eq.~(%
\ref{y(t)})], completes the model of qubit-based systems as realizations of
memory circuit elements.

\section{Illustrative examples}

\label{sec4}

Frequency-dependent pinched hysteresis loops are the most pronounced
signatures of memory circuit elements~[%
\onlinecite{chua76a,diventra09a,pershin11a}]. In this section we consider
qubit-based memcapacitors and meminductors subjected to a periodic input, $%
\varepsilon _{1}(t)\propto u(t)=u_{0}\sin \omega t$. The examples presented
below highlight the unusual dynamical features of these quantum devices.

We emphasize that Figs.~\ref{Fig:Rabi}-\ref{Fig:Delayed} illustrate
hysteresis curves for both charge and flux qubits. For the charge qubit: $%
u=V $, $u_{0}=V_{\mathrm{A}}$, $y=Q$, and $y_{0}=C_{\mathrm{geom}}V_{\mathrm{%
A}}$. For the flux qubit: $u=I$, $u_{0}=I_{\mathrm{A}}$, $y=\phi $, and $%
y_{0}=LI_{\mathrm{A}}$.

\subsection{Rabi oscillations}

\label{sec:Rabi}

Consider the situation when the applied frequency is close to the resonance
frequency, so that $\delta \omega \equiv \omega -\omega _{\mathrm{q}}\ll
\omega $. If the relaxation time $\Gamma ^{-1}$ is quite long, one can
ignore the relaxation and find an analytical solution for the problem (see
Appendix B for details). In particular, precisely at the resonance ($\omega
=\omega _{\mathrm{q}}$), we obtain:%
\begin{equation}
\left\langle \sigma _{z}\right\rangle =\frac{\varepsilon _{0}}{\Delta E}\cos
\Omega _{\mathrm{R}}^{(0)}t-\frac{\Delta }{\Delta E}\sin \Omega _{\mathrm{R}%
}^{(0)}t\cos \omega t\text{,}  \label{sigma_Z}
\end{equation}%
where $\Omega _{\mathrm{R}}^{(0)}$ is given by Eq. (\ref{OR}). Equation~(\ref%
{sigma_Z}), describing the Rabi oscillations, can be further simplified, at
both the avoided-level crossing and far from this point:%
\begin{eqnarray}
\varepsilon _{0} &=&0:\quad \quad \,\left\langle \sigma _{z}\right\rangle
=-\sin \Omega _{\mathrm{R}}^{(0)}t\cos \omega t\text{,}  \label{sigma_Z1} \\
\left\vert \varepsilon _{0}\right\vert &\gg &\Delta :\quad \quad
\left\langle \sigma _{z}\right\rangle =\mathrm{sign}(\varepsilon _{0})\cos
\Omega _{\mathrm{R}}^{(0)}t.
\end{eqnarray}

\begin{figure}[th]
\includegraphics[width=8cm]{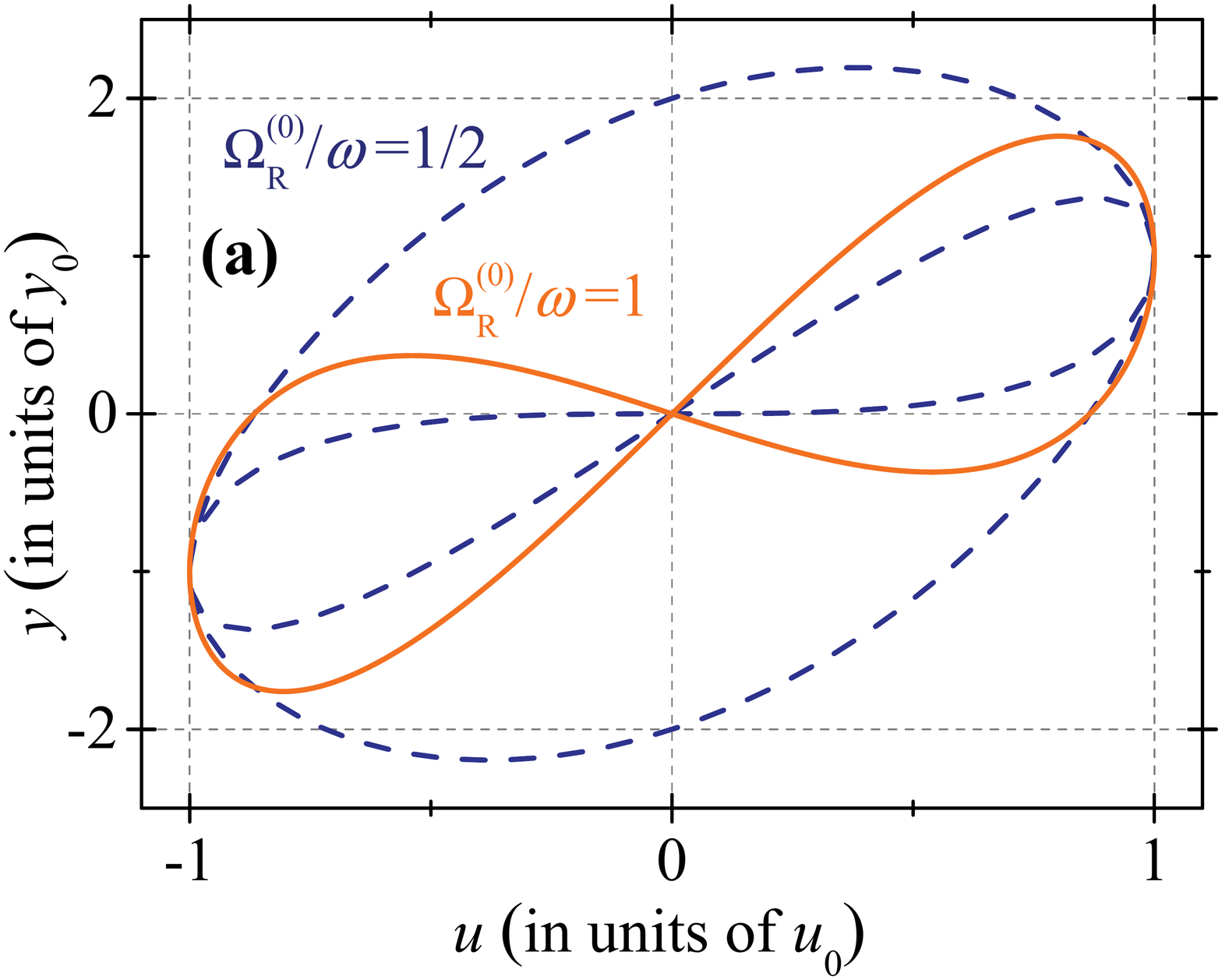} \includegraphics[width=8cm]{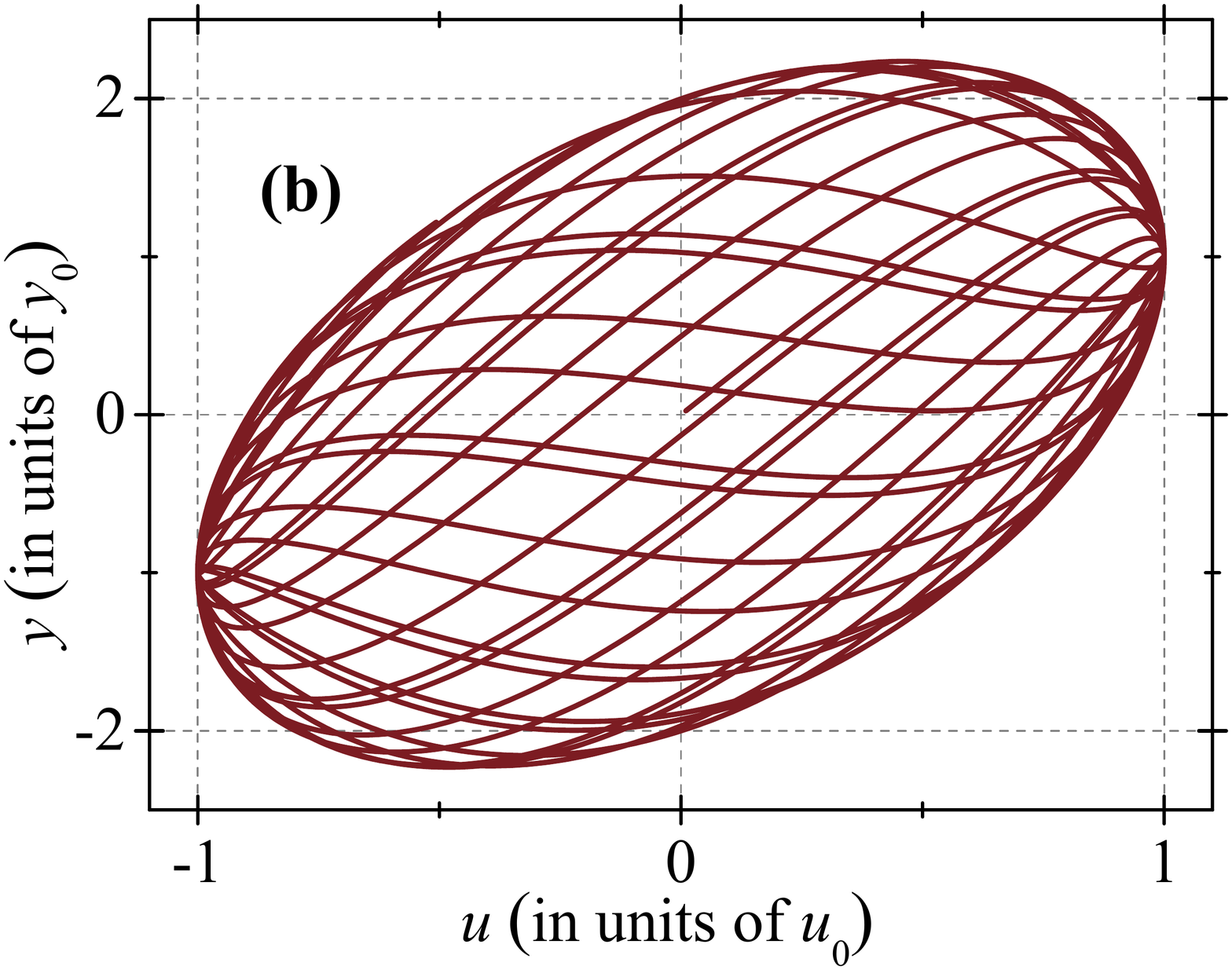}
\caption{$y$ versus $u$ hysteresis curves in the Rabi
oscillation regime for (a) commensurate and (b) incommensurate frequencies.
This plot was obtained using Eq.~(\protect\ref{phi/phi0}) and the following
set of parameter values: $\protect\omega =\protect\omega _{\mathrm{q}}$, $%
\varkappa =2$, $\Omega _{\mathrm{R}}^{(0)}/\protect\omega =1$ (solid curve
in (a)), $\Omega _{\mathrm{R}}^{(0)}/\protect\omega =1/2$ (dashed curve in
(a)), $\Omega _{\mathrm{R}}^{(0)}/\protect\omega =1/\protect\sqrt{2}$ (b).
(b) presents the curve corresponding to about $16$ periods of the sinusoidal
input.}
\label{Fig:Rabi}
\end{figure}

The case of $\varepsilon _{0}=0$ and $\delta \omega =0$ corresponds to the
excitation by $u(t)=u_{0}\sin \omega _{\mathrm{q}}t$. Using Eq.~(\ref{u(t)})
we find the system response in this case:
\begin{equation}
\frac{y(t)}{y_{0}}=\sin \omega t+\varkappa \sin \Omega _{\mathrm{R}%
}^{(0)}t\cos \omega t.  \label{phi/phi0}
\end{equation}%
The shape of the hysteresis curve (\ref{phi/phi0}) is defined by the
commensurability of $\omega $ and $\Omega _{\mathrm{R}}^{(0)}$. In
particular, if the ratio of these frequencies is a rational number, $\Omega
_{\mathrm{R}}^{(0)}/\omega =n/m$ (here, $n$ and $m$ are integers), then the
hysteresis curve is a closed loop. One can show that the period of such a
loop is $T^{\ast }=nT_{\mathrm{R}}=mT_{\omega }$, where $T_{\mathrm{R}}=2\pi
/\Omega _{\mathrm{R}}^{(0)}$ and $T_{\omega }=2\pi /\omega $ are periods of
the Rabi oscillations and periodic input, respectively. In the opposite case
of an irrational ratio $\Omega _{\mathrm{R}}^{(0)}/\omega $, the curve is
not closed. Both cases are illustrated in Fig.~\ref{Fig:Rabi}.

We note that Fig.~\ref{Fig:Rabi} was obtained using Eq.~(\ref{phi/phi0})
found in the rotating-wave approximation (see Appendix B). However, at high
driving amplitudes, the resonant frequency is shifted according to the
Bloch-Siegert expression~[%
\onlinecite{Bloch40a,Shevchenko08,Tuorila10a,Romhanyi15a}]. The corrected
resonant frequency can be found numerically by solving the Liouville
equation. For the selected set of parameter values, an \textquotedblleft $8$%
-shaped\textquotedblright\ closed hysteresis loop [as in Fig. \ref{Fig:Rabi}%
(a)] is obtained for $\Omega _{\mathrm{R}}^{(0)}/\omega =1.045$, instead of $%
\Omega _{\mathrm{R}}^{(0)}/\omega =1$ as predicted in the rotating-wave
approximation.

An important feature of the qubit-based memory devices is that their
characteristic operational frequencies $\omega _{\mathrm{q}}$\ belong to the
gigahertz region. Such frequencies make the devices controllable by
microwaves, on short timescales.

\subsection{Two-photon excitation}

\label{2ph_exc}

Next, we consider a different excitation regime, when the driving frequency
is at half the qubit frequency, $\omega =\omega _{\mathrm{q}}/2$. This is
the two-photon process~[\onlinecite{Shevchenko12}] as two photons are
required to excite the qubit. The two-photon process is characterized by its
own Rabi frequency (see Ref.~[\onlinecite{Shevchenko14}] for more details)
that, together with the excitation frequency $\omega $ (and possibly some
other frequencies) defines the system response.

\begin{figure}[h]
\includegraphics[width=8cm]{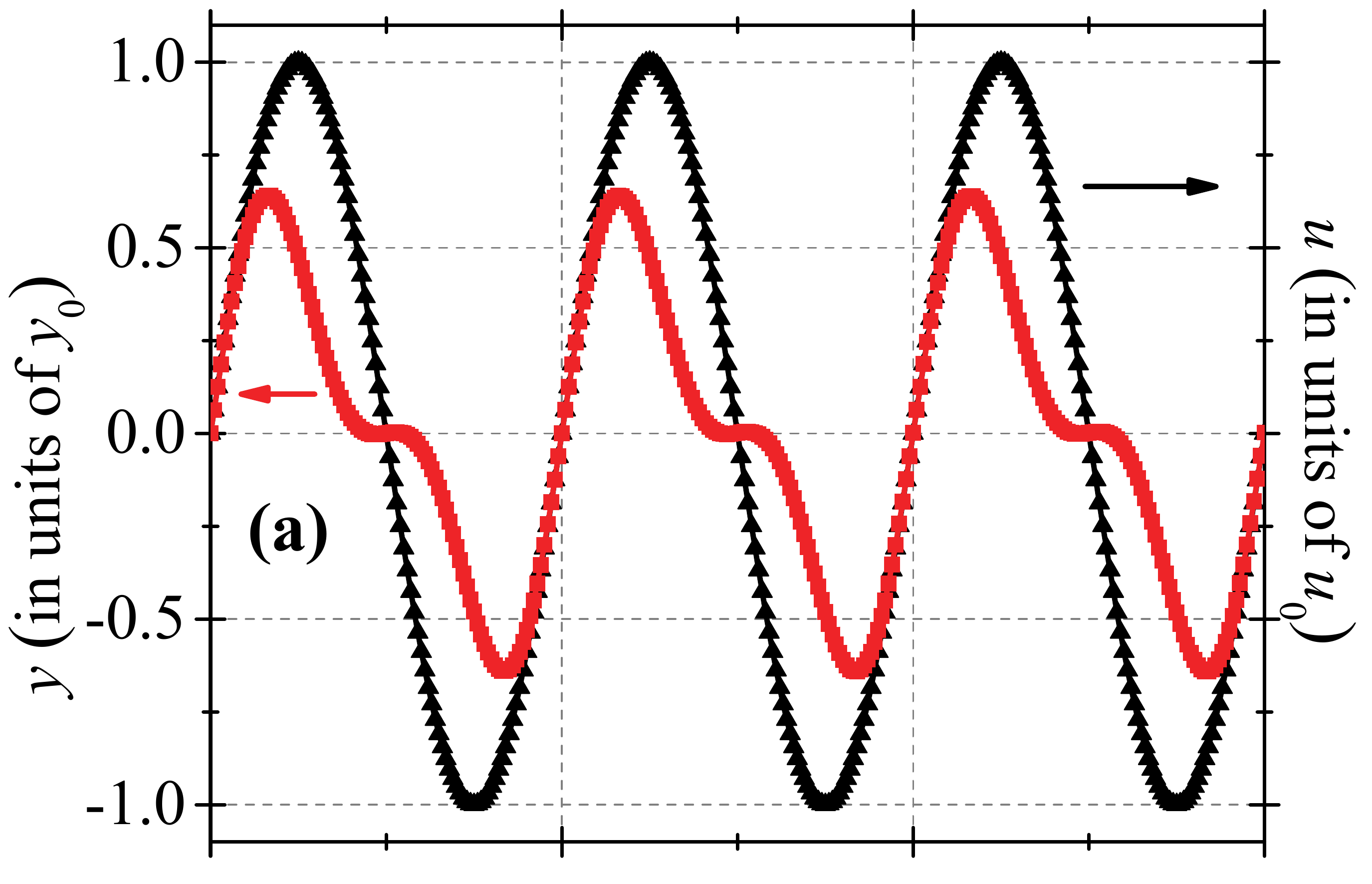} \includegraphics[width=8cm]{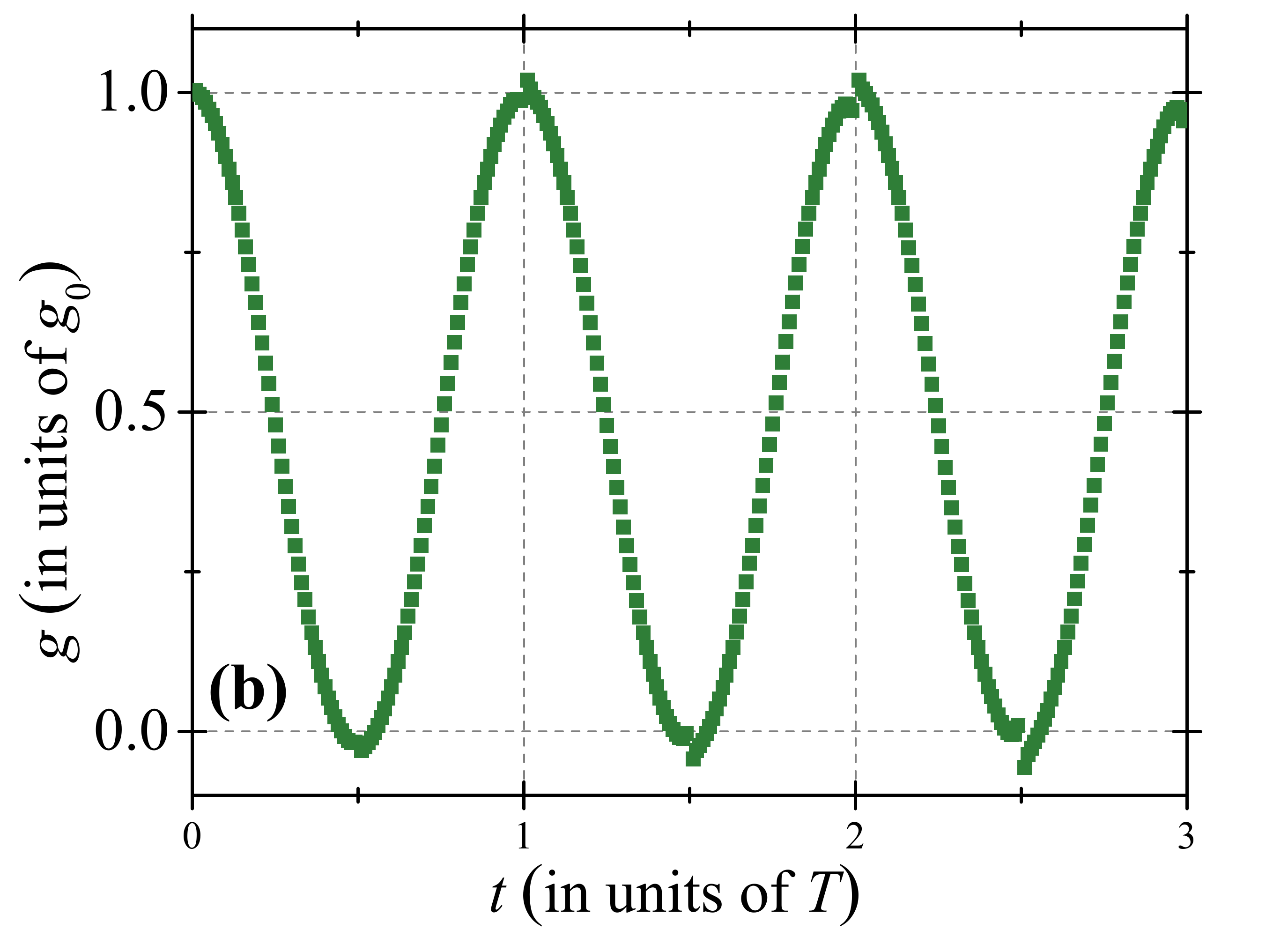} %
\includegraphics[width=7.5cm]{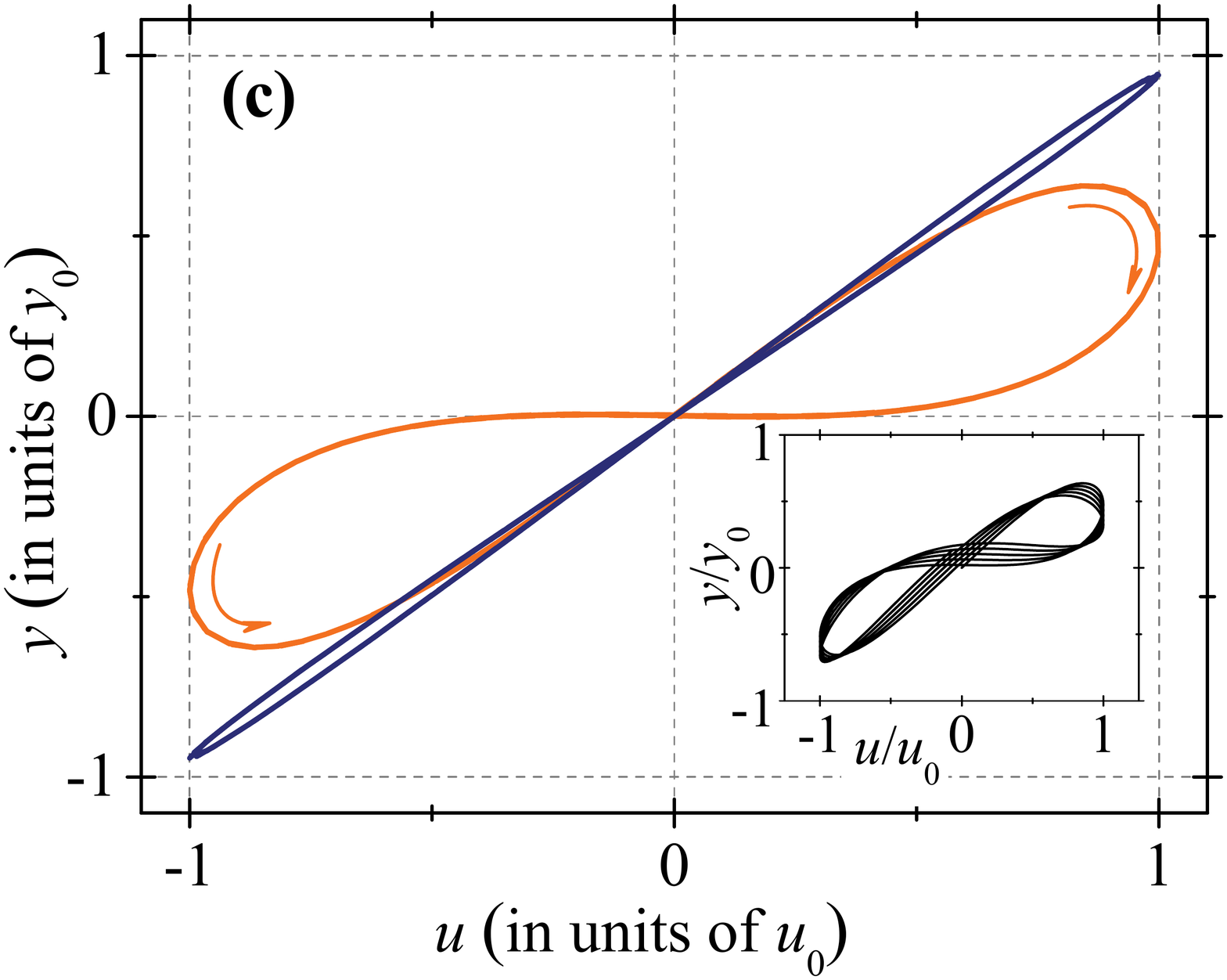}
\caption{Time-dependencies of (a) the input $u$ and output $%
y $, and (b) the generalized response $g$ for the case of two-photon
excitation. (c) $y$ versus $u$ hysteresis curves. These plots were obtained
using the following parameter values: $\protect\omega =1.013\cdot \protect%
\omega _{\mathrm{q}}/2$, $\protect\varepsilon _{0}=0$, $\varkappa =2$, $%
\Gamma =0$, and $A=0.2\Delta $. The narrow hysteresis curve in (c) is found
at a different value of $\varkappa =0.2$. The inset shows the hysteresis
curve found at a different value of $\protect\omega =\protect\omega _{%
\mathrm{q}}/2$. The inset curve corresponds to five periods of the input
oscillations.}
\label{Fig:two-photon}
\end{figure}

In particular, at zero offset $\varepsilon _{0}=0$, the two-photon Rabi
frequency is zero~[\onlinecite{Shevchenko14}]. The excitation at the
two-photon resonant frequency accounted for the Bloch-Siegert shift ($\omega
=1.013\cdot \omega _{\mathrm{q}}/2$) results in a closed \textquotedblleft $%
8 $-shaped\textquotedblright\ hysteretic curve depicted in Fig. \ref%
{Fig:two-photon}(c), demonstrating the periodicity defined by $\omega $.
Figure~\ref{Fig:two-photon}(a)-(b) shows the time dependencies of the input,
output, and generalized response function (plotted in units of $%
g_{0}=y_{0}/u_{0}$) found in the same calculation by solving the Bloch
equation. The shift in the excitation frequency from the resonant one
introduces a new periodicity in the response, as shown in the inset of Fig. %
\ref{Fig:two-photon}(c). The origin of this modification most probably could
be related to a shift of the two-photon Rabi frequency from zero.

\subsection{Delayed response}

\label{del_resp}

In the previous subsections, we considered the resonant excitations in which
the driving frequency is an integer number of the qubit frequency $\omega _{%
\mathrm{q}}$, and $\varepsilon _{0}=0$. Here, we consider the opposite case,
when the excitation frequency is small and far from the resonance. In this
situation, the qubit demonstrates a lag, the finite time needed for the
qubit to come into equilibrium. Because of large detuning, there are no Rabi
oscillations in this regime.

\begin{figure}[th]
\includegraphics[width=8.7cm]{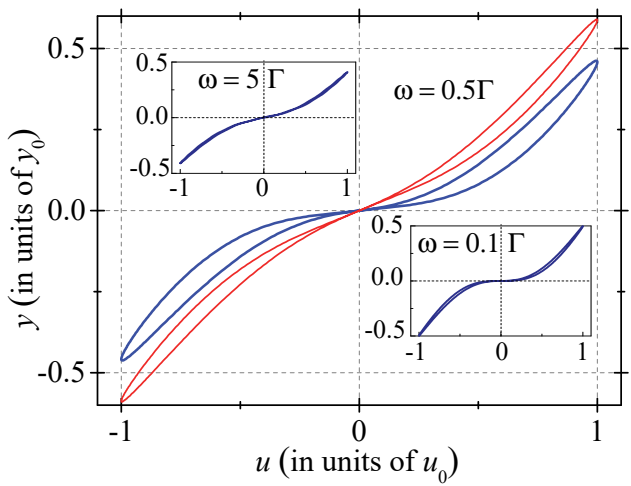}
\caption{$y$ versus $u$ hysteretic curves in the delayed
response regime. These curves were obtained using the following parameter
values: $\Gamma =0.01\Delta /\hbar $, $A=\Delta $, $\varkappa =1$, and $%
\protect\varepsilon _{0}=0$. The values of $\protect\omega $ are indicated
on the plot. All curves are calculated for zero temperature, except of the
thin red curve, which illustrates the effect of the temperature, with $%
T=0.5\Delta /k_{\mathrm{B}}$.}
\label{Fig:Delayed}
\end{figure}

In order to demonstrate the lagging effect, we select the driving frequency $%
\omega $ comparable to the relaxation rate, for which we choose $\Gamma
=0.01\Delta $, and solve the Bloch equations (\ref{eq4x}) numerically.
Figure~\ref{Fig:Delayed} shows selected results of these calculations. We
found that the largest size hysteresis is observed when the input frequency $%
\omega $ is of the order of $\Gamma /2$. This feature was discussed in
detail in Refs.~[%
\onlinecite{Grajcar08, Nori08, Gonzalez15, Shevchenko15,
Okazaki16}] %.

The insets in Fig.~\ref{Fig:Delayed} demonstrate that at lower and higher
driving frequencies (compared to $\Gamma $) the hysteresis vanishes. Indeed,
if the frequency is high, the system does not have enough time to relax to
equilibrium. At very low frequencies, the system stays very close to the
equilibrium at every instant of time, so that the hysteresis is not observed.

Let us finally discuss the effect of the temperature, which was ignored so
far. The temperature enters the Bloch equation through the factor $%
Z_{0}=\tanh (\Delta E/2k_{\mathrm{B}}T)$ as well as through the temperature
dependence of the decoherence and relaxation rates. Neglecting the latter,
the red thin line in Fig.~\ref{Fig:Delayed} illustrates that a significant
temperature, $T\sim \Delta /k_{\mathrm{B}}$, changes the shape of the curve
and reduces the width of the hysteresis loop. At $T\gg \Delta /k_{\mathrm{B}%
} $, this width tends to zero and the dependence becomes linear. So, in
order to ignore the temperature in this context, it should be much smaller
than $\Delta /k_{\mathrm{B}}$, which is usually the case in the experimental
realizations of qubit-based systems.

\subsection{General picture}

\label{general}

\begin{figure}[th]
\includegraphics[width=8cm]{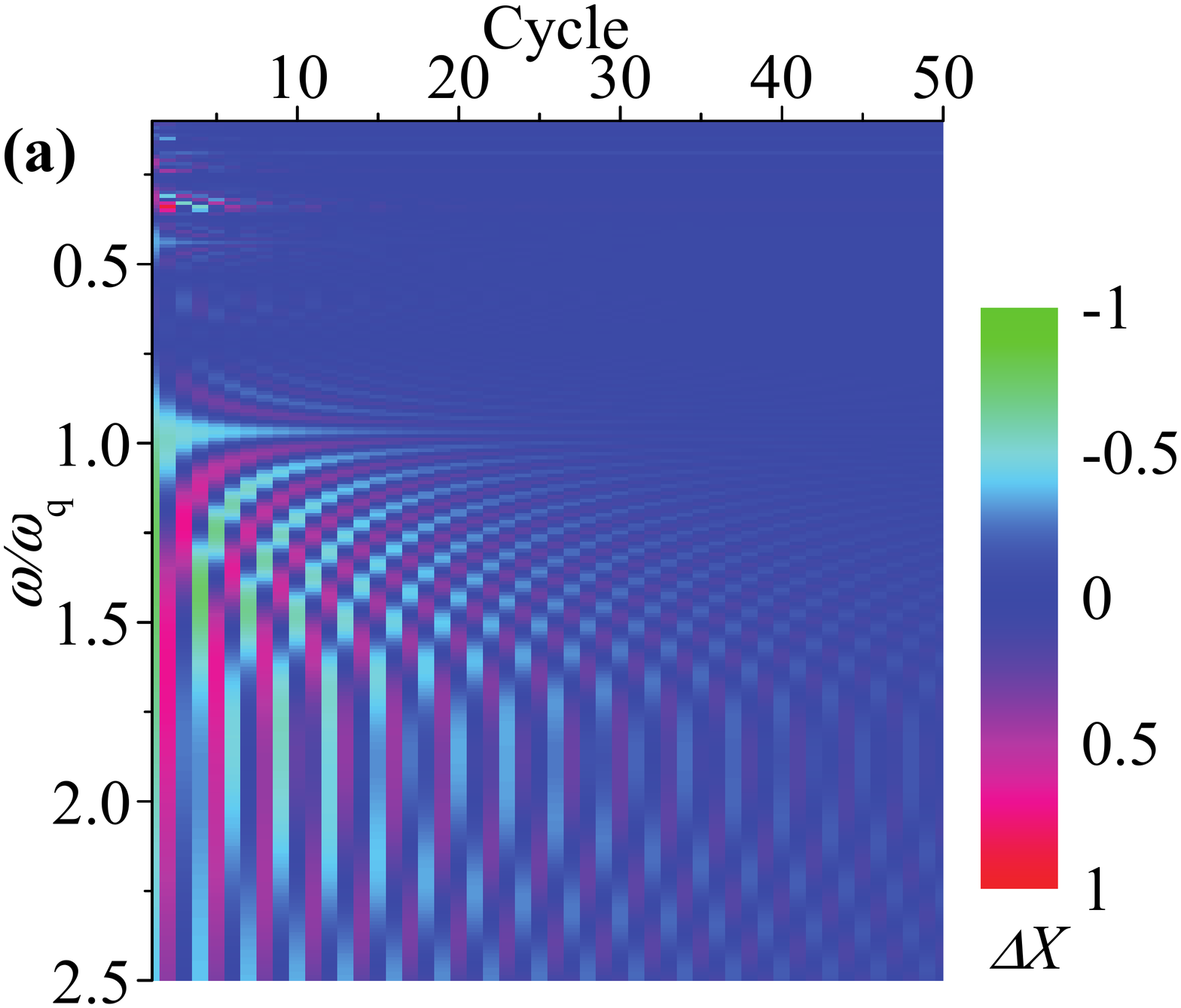} \includegraphics[width=8cm]{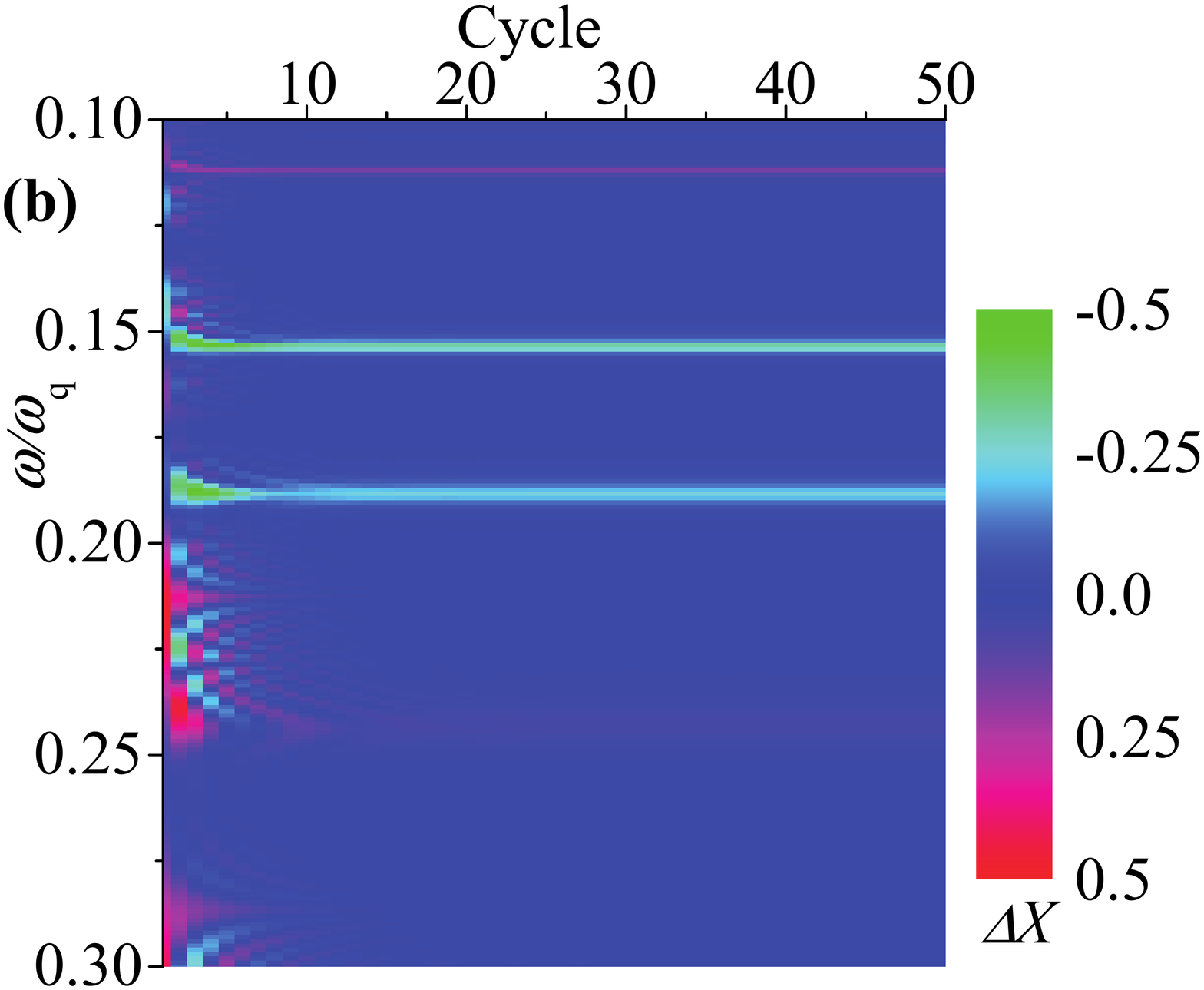}
\newline
\caption{(a) Hysteresis of $X$ as a function of $\protect%
\omega $ and cycle calculated using Eq.~(\protect\ref{DX}). This plot was
obtained employing the following set of parameters: $\protect\varepsilon %
_{0}=0$, $\Omega _{\mathrm{R}}^{(0)}/\protect\omega _{\mathrm{q}}=1$, and $%
\Gamma =0.01\Delta /\hbar $. (b) Zoom of the low-frequency region in (a).}
\label{Fig:map1}
\end{figure}

In order to better understand the features of the hysteretic response, we
plot the hysteresis of $X$ in different cycles of the sinusoidal input
calculated as
\begin{equation}
\Delta X_{n}=X(3T_{\omega }/8+nT_{\omega })-X(T_{\omega }/8+nT_{\omega }),
\label{DX}
\end{equation}%
where $n$ is the number of the cycle, and $T_{\omega }=2\pi /\omega $ is the
period of the input signal. Figure \ref{Fig:map1} presents an example of
such calculation. In particular, in Fig. \ref{Fig:map1}(a) at $\omega
/\omega _{\mathrm{q}}$ slightly less than unity, one can clearly recognize
the Rabi oscillations corresponding to the solid line in Fig. \ref{Fig:Rabi}%
(a). At lower frequencies [see \ref{Fig:map1}(b)] one can distinguish
several horizontal lines of a fixed-size hysteresis. These likely correspond
to the $k$-photon processes at $\omega /\omega _{\mathrm{q}}\approx 1/k$,
similarly to the two-photon case with $k=2$ considered above.

Few other plots of hysteresis in $X$ found at different parameter values are
provided in Appendix C.

\section{Conclusion}

\label{sec5}

We demonstrated that several qubit-based structures belong to the class of
memory circuit elements~[\onlinecite{diventra09a}]. It was shown that when
subjected to a periodic input, such qubit-based memcapacitive and
meminductive systems exhibit frequency-dependent hysteresis curves. Note
that the quantumness of superconducting qubits requires special care in
performing and interpreting experiments with such devices. In particular, in
addition to the basic components considered in this work, realistic
experimental setups include an additional apparatus for measuring the
quantum subsystem state. For the sake of simplicity, this issue was not
addressed here, since our aim is to demonstrate the aspects of the qubit
dynamics relevant to memory devices. Various measurement techniques are
discussed, e.g., in Ref.~[\onlinecite{Wendin07}].

Our work not only extends the set of memory circuit elements with novel
components featuring an unusual and rich quantum dynamics of their internal
states, but may also result in novel applications of qubit-based structures
beyond the ones traditionally considered for superconducting qubits.

\begin{acknowledgments}
S.N.S. is grateful to A. Fedorov, M.F. Gonzalez-Zalba, A.N. Omelyanchouk,
and O.G. Turutanov for valuable discussions. This work has been supported by
the NSF grant No. ECCS-1202383, USC Smart State Center for Experimental
Nanoscale Physics, RIKEN iTHES Project, MURI Center for Dynamic
Magneto-Optics, a Grant-in-Aid for Scientific Research (A), a grant from the
John Templeton Foundation, the State Fund for Fundamental Research of
Ukraine, and the Russian Scientific Foundation grant No. 15-13-20021.
\end{acknowledgments}

%\bibliographystyle{plain}

%\bibliography{memcapacitor}

%\clearpage

%\setcounter{equation}{0} \renewcommand{\theequation}{S\arabic{equation}} %
%\setcounter{figure}{0} \renewcommand{\thefigure}{S\arabic{figure}} %
%\setcounter{page}{1}

%\section*{Supplementary Material for "Qubit-based memcapacitors and meminductors"}

\appendix

\section{Meminductance of a Josephson junction}

\begin{figure}[t]
\includegraphics[width=8.7cm]{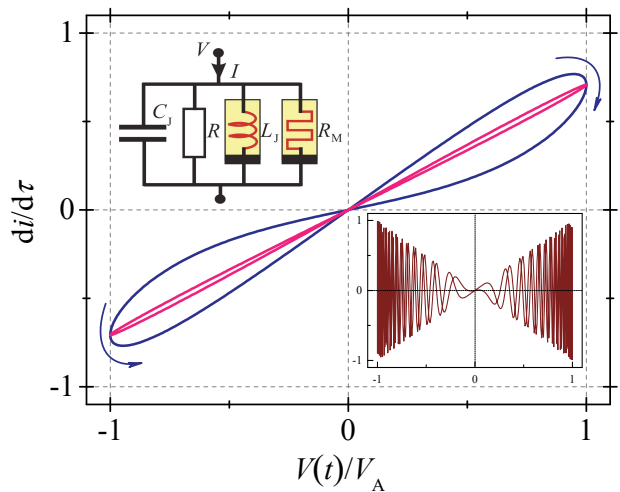}
\caption{Josephson junction as a meminductor. Left upper
inset: model of a Josephson junction which describes the Josephson
inductance $L_{\mathrm{J}}$ as the meminductance and also includes the
memristance $R_{\mathrm{M}}$. Main panel: dependence of the current
time-derivative on the applied voltage. The voltage $V(t)$ is normalized by
its amplitude $V_{\mathrm{A}}$; the dimensionless current is $i=I/I_{\mathrm{%
c}}$ and time $\protect\tau =\frac{2eV_{\mathrm{A}}}{\hbar }t$. The thick
and thin lines are plotted for $\frac{\hbar \protect\omega }{2eV_{\mathrm{A}}%
}=2$ and $20$, respectively, while the right bottom inset is plotted for
this value being $0.01$. For all three curves we have taken $\protect\varphi %
_{0}=\protect\pi /4$.}
\label{Fig:JJ}
\end{figure}

Here we present several results which allow interpreting a Josephson
junction as a memory device. From the original work by Josephson, it is
known that the resistance of the junction contains a phase-dependent term~[%
\onlinecite{Josephson62, Likharev}], which can be treated as a memristance~[%
\onlinecite{chua03a}]. This was recently studied in Ref.~%
\onlinecite{Peotta14}. In addition to this, the well-known Josephson
inductance can be treated as a meminductance. From this perspective,
according to Ref.~\onlinecite{chua03a}, the correct model for a Josephson
junction should include a resistor $R$, a capacitor $C_{\mathrm{J}}$, a
memristor $R_{\mathrm{M}}$, and a meminductor $L_{\mathrm{J}}$, as shown in
Fig.~\ref{Fig:JJ}, left upper inset. These aspects deserve special
attention. So, before considering Josephson-junction-based effective
two-level systems, qubits, let us describe here the Josephson meminductance.
For simplicity, we will not address here neither the phase-dependent
memristance, nor other aspects, which result in hysteretic dependencies.
Note that the phase-dependent memristance in a related context was studied
both in the classical [\onlinecite{Peotta14}] and quantum [%
\onlinecite{Salmilehto16}] regimes.

It is known that a Josephson junction can be described as a Josephson
inductance. This directly follows from the two Josephson relations, which
relate the current $I$ and the voltage $V$ with the order parameter phase
difference $\varphi $:%
\begin{eqnarray}
I(t) &=&I_{\mathrm{c}}\sin \varphi (t),  \label{J1} \\
V(t) &=&\frac{\Phi _{0}}{2\pi }\dot{\varphi},  \label{J2}
\end{eqnarray}%
where $I_{\mathrm{c}}$\ is the critical current of the junction and $\Phi
_{0}=h/2e$ is the flux quantum. These can be rewritten as%
\begin{equation}
V(t)=\frac{\Phi _{0}}{2\pi \cos \varphi }\dot{I}\equiv L_{\mathrm{J}%
}(\varphi )\dot{I},
\end{equation}%
so that the proportionality term is used for the definition of the
inductance $L_{\mathrm{J}}$. This inductance is often referred to as the
\textit{nonlinear inductance}. Strictly speaking, to be nonlinear, this must
be a function of the voltage $V$, namely it must be determined by the
instantaneous value of the voltage. Instead, the inductance depends on the
phase difference $\varphi $, defined by the voltage history. That is why,
following Ref.~\onlinecite{chua03a}, we argue that it would be more correct
to call this a \textit{memory inductance}, or meminductance. Indeed,
integrating Eq.~(\ref{J2}), one obtains%
\begin{equation}
\varphi (t)=\varphi _{0}+\frac{\Phi _{0}}{2\pi }\int\limits_{0}^{t}V(t^{%
\prime })dt^{\prime }.  \label{phi}
\end{equation}%
Consequently, the inductance of the Josephson junction $L_{\mathrm{J}%
}(\varphi )$ is precisely the memory inductance, since the phase $\varphi $
has a memory of the voltages applied in the past.

One can introduce the generalized flux as%
\begin{equation}
\Phi =\int\limits_{0}^{t}V(t^{\prime })dt^{\prime },  \label{Flux}
\end{equation}%
which relates it to the phase difference $\varphi $ in Eq.~(\ref{phi}). Then
we can rewrite the above expressions, so that the Josephson junction is
obviously a \textit{flux-controlled meminductive system} with the control
parameter $x=2\pi \frac{\Phi }{\Phi _{0}}$:%
\begin{eqnarray}
\dot{I} &=&L_{\mathrm{J}}^{-1}(\Phi )V,  \label{Idot} \\
\dot{\Phi} &=&V.  \label{Fidot}
\end{eqnarray}%
We note that for a non-linear element there are different possibilities to
introduce the inductance [\onlinecite{Likharev}], and equation~(\ref{Idot})
is one possibility. For an alternative definition see Eq.~(\ref{LM}), which
is obtained from Eq.~(\ref{Idot}); then $L_{\mathrm{M}}=\int \dot{I}L_{%
\mathrm{J}}\mathrm{d}t/I$.

To further explore these relations, let us now consider a junction biased by
the alternating voltage $V(t)=V_{\mathrm{A}}\cos \omega t$. In Fig.~\ref%
{Fig:JJ} we plot the dependence of the current derivative on the voltage.
This hysteretic dependence was plotted for three values of the frequency.
This is shown in the thin-line narrow hysteresis for high frequency; the
optimal hysteretic loop for the intermediate case, shown by the thick line;
and the complicated hysteretic curve for the low frequency, which is
presented in the right bottom inset.

\subsection{Hysteresis in the CRSJ model}

Consider now the CRSJ model as described above, but with the resistance $R$
and capacitance $C_{\mathrm{J}}$ taken into account explicitly. For
definiteness, consider the voltage-biased regime with the applied voltage $%
V(t)=V_{\mathrm{A}}\cos \omega t$. The current is given by the extended
version of Eq.~(\ref{J1}),%
\begin{equation}
I(t)=I_{\mathrm{c}}\sin \varphi (t)+\frac{V(t)}{R}+C_{\mathrm{J}}\dot{V}(t),
\label{J1b}
\end{equation}%
which reflects the Kirchhoff law for the circuit shown in Fig.~\ref{Fig:JJ}.
There, for simplicity, we disregard the memristance, to accentuate on the
meminductance. With $\varphi (t)$ given by Eq.~(\ref{phi}), the
dimensionless version of Eq.~(\ref{J1b}) reads:%
\begin{equation}
\frac{di}{d\tau }=\cos \!\left( \!\varphi _{0}+\frac{\sin w\tau }{w}%
\!\!\right) \cos w\tau -\frac{\Omega }{Q}\sin w\tau -\Omega ^{2}\cos w\tau .
\label{di/dtau}
\end{equation}%
Here we introduced the dimensionless values
\begin{equation}
i=\frac{I}{I_{\mathrm{c}}}\text{, \ }w=\frac{\hbar \omega }{2eV_{\mathrm{A}}}%
\text{, \ }\tau =\frac{\omega t}{w}\text{, \ }\Omega =\frac{\omega }{\omega
_{\mathrm{p}}},
\end{equation}%
with the Josephson plasma frequency $\omega _{\mathrm{p}}$ and the quality
factor $Q$ defined by the capacitance $C_{\mathrm{J}}$ and the resistance $R$%
, respectively, as follows%
\begin{equation}
\omega _{\mathrm{p}}^{2}=\frac{2eI_{\mathrm{c}}}{\hbar C_{\mathrm{J}}}=\frac{%
2E_{\mathrm{C}}E_{\mathrm{J}}}{\hbar ^{2}}\text{, \ }Q^{2}\equiv \beta =%
\frac{2e}{\hbar }I_{\mathrm{c}}R^{2}C_{\mathrm{J}}.
\end{equation}%
Here the plasma frequency is also expressed with the characteristic
charging, $E_{\mathrm{C}}=(2e)^{2}/2C_{\mathrm{J}}$, and Josephson, $E_{%
\mathrm{J}}=\hbar I_{\mathrm{c}}/2e$, energies of the contact, and $\beta $
is the Stewart-McCumber parameter.

\begin{figure}[th]
\includegraphics[width=8.7cm]{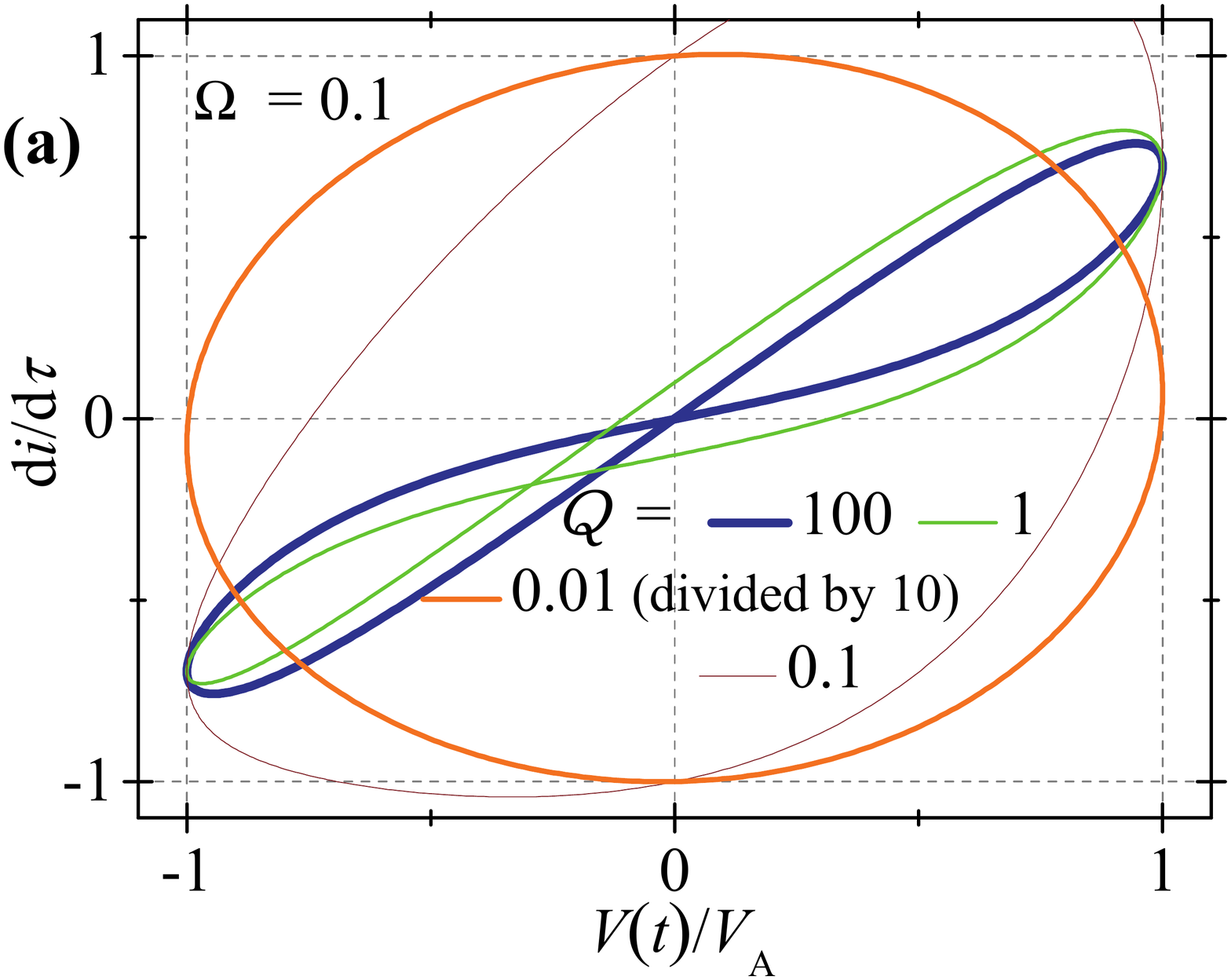} %
\includegraphics[width=8.7cm]{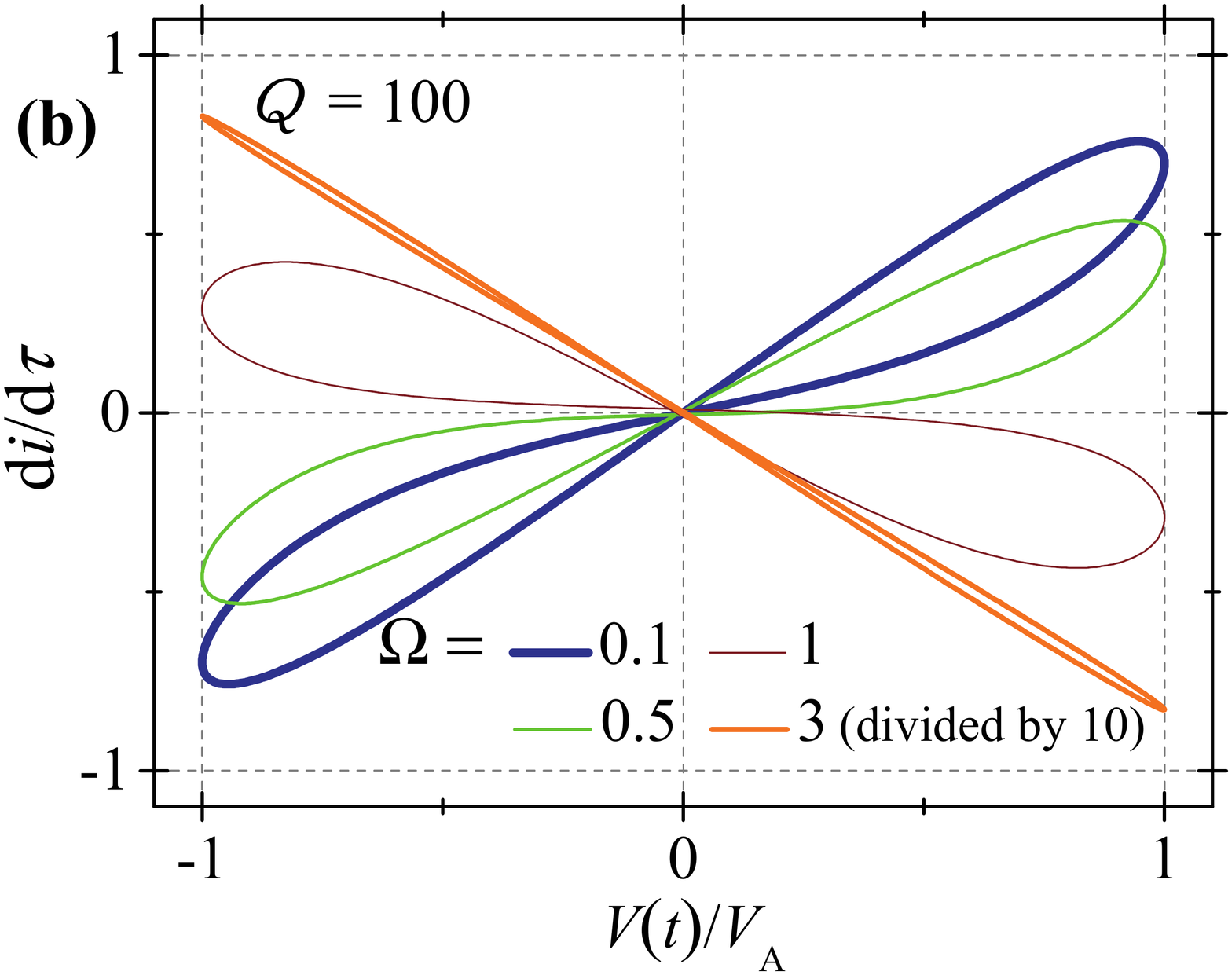}
\caption{Impact of the resistance $R$ (upper panel) and
capacitance (bottom panel): Dependence of the current time-derivative on the
applied voltage. The dimensionless current is $i=I/I_{\mathrm{c}}$ and the
reduced time $\protect\tau =\protect\omega t/w$. The parameters used here
are: $w=2$ and $\protect\varphi _{0}=\protect\pi /4$; the resistance (i.e., $%
Q$) and capacitance (i.e., $\Omega =\protect\omega /\protect\omega _{\mathrm{%
p}}$) are varied, as shown in the legends.}
\label{Fig:JJ_with_RC}
\end{figure}

In Fig.~\ref{Fig:JJ_with_RC} we explore the impact of the resistance $R$ and
the capacitance $C_{\mathrm{J}}$ on the hysteresis considered previously in
Fig.~\ref{Fig:JJ}; note that Fig.~\ref{Fig:JJ} corresponds to $Q\rightarrow
\infty $ and $\Omega \rightarrow 0$. Our numerical calculations demonstrate
that there is a pinched hysteresis loop for $Q\gtrsim 100$ and $\Omega
\lesssim 0.1$. For realistic junctions, it seems that there is no problem
both with the former condition of weak damping ($\beta \gg 1$) and with the
latter condition ($\omega \ll \omega _{\mathrm{p}}$) of neglecting the
displacement current next to the Josephson one~[\onlinecite{Likharev}].

\section{Dynamics of the two-level system}

Consider now the dynamics of a two-level system, which we consider here, for
clarity, for the meminductive case with the flux qubit. The generalization
to other cases, such as the one of the charge qubit, is obvious.

The current in the qubit loop is defined~[\onlinecite{Makhlin02, Wendin07}]
by its operator, given by $-I_{\mathrm{p}}\sigma _{z}$. In order to take
into account relaxation processes, one has to consider the energy
representation. Let the qubit density matrix in this representation be
parameterized as follows: $\rho =\frac{1}{2}(1+\mathbf{x\sigma })$. Changing
from the flux representation to the energy representation is executed by
means of the matrix $S=\left(
\begin{array}{cc}
\cos \zeta /2 & \sin \zeta /2 \\
-\sin \zeta /2 & \cos \zeta /2%
\end{array}%
\right) $ with $\tan \zeta =-\Delta /\varepsilon _{0}$. Then the qubit
current becomes
\begin{equation}
I_{\mathrm{q}}=-I_{\mathrm{p}}\left\langle \sigma _{z}\right\rangle =I_{%
\mathrm{p}}\left( \frac{\Delta }{\Delta E}X-\frac{\varepsilon _{0}}{\Delta E}%
Z\right) .  \label{Iq}
\end{equation}%
One can see that in the \textit{ground/excited state}, $X=0$ and $Z=\pm 1$: $%
I_{\mathrm{q}}=\mp I_{\mathrm{p}}\frac{\varepsilon _{0}}{\Delta E}$, with
zero current at the avoided level crossing, for $\varepsilon _{0}=0$, and
with $I_{\mathrm{q}}=\pm I_{\mathrm{p}}$ far from it, at $\left\vert
\varepsilon _{0}\right\vert \gg \Delta $.

The qubit current in Eq.~(\ref{Iq}) is defined by the difference between the
probabilities of the currents in the two directions $\left\langle \sigma
_{z}\right\rangle $, which is calculated by solving$\ $the Bloch equation [%
\onlinecite{Blum, Makhlin02, Wendin07}]:%
\begin{eqnarray}
\dot{X} &=&-B_{z}Y-\Gamma _{2}X,  \label{Bloch} \\
\dot{Y} &=&B_{z}X-B_{x}Z-\Gamma _{2}Y,  \notag \\
\dot{Z} &=&B_{x}Y-\Gamma _{1}(Z-Z_{0}).  \notag
\end{eqnarray}%
Here $\Gamma _{1,2}=T_{1,2}^{-1}$ is the energy and phase relaxation rates, $%
Z_{0}=\tanh (\Delta E/2k_{\mathrm{B}}T)$ corresponds to the equilibrium
energy level populations, and%
\begin{eqnarray}
B_{x}\!\! &=&\!\!\frac{\Delta }{\Delta E}\frac{\varepsilon _{1}(t)}{\hbar }%
\equiv 2\Omega _{\mathrm{R}}^{(0)}\sin \omega t,\text{\ \ }\Omega _{\mathrm{R%
}}^{(0)}\!=\!\frac{\Delta A}{2\hbar \Delta E}, \\
B_{z}\!\! &=&\!\!-\omega _{\mathrm{q}}-\frac{\varepsilon _{0}}{\Delta }B_{x}.
\end{eqnarray}

These equations can also be written in vector form (to better correspond to
the theory of memory-devices~[\onlinecite{pershin11a, diventra09a}]):%
\begin{equation}
\mathbf{\dot{x}}\!=\!\mathbf{f}(\mathbf{x},I)\!\equiv \!\mathbf{B\times x}%
-\Gamma _{2}\mathbf{x}_{\Vert }\mathbf{-}\Gamma _{1}\mathbf{(x}_{\bot }-%
\mathbf{x}_{0}),  \label{Bloch_vector}
\end{equation}%
\begin{equation}
\mathbf{x}\!\mathbf{=}\!\left(
\begin{array}{c}
X \\
Y \\
Z%
\end{array}%
\right) \!\!,\text{ }\mathbf{B\!=}\!\!\left(
\begin{array}{c}
B_{x} \\
0 \\
B_{z}%
\end{array}%
\right) \!\!,\text{\ }\mathbf{x}_{0}\mathbf{=}\!\left(
\begin{array}{c}
0 \\
0 \\
Z_{0}%
\end{array}%
\right) \!\!,
\end{equation}%
\begin{equation}
B_{x}(I)\!\!=\!\!\frac{\Delta }{\Delta E}\frac{2MI_{\mathrm{p}}}{\hbar }I,%
\text{\ }B_{z}(I)\!\!=\!\!-\omega _{\mathrm{q}}\!-\!\frac{\varepsilon _{0}}{%
\Delta E}\frac{2MI_{\mathrm{p}}}{\hbar }I.
\end{equation}%
Here the longitudinal and transversal components of the vector are given by $%
\mathbf{x}_{\Vert }=(\mathbf{xe}_{x})\mathbf{e}_{x}+(\mathbf{xe}_{y})\mathbf{%
e}_{y}$ and $\mathbf{x}_{\bot }=(\mathbf{xe}_{z})\mathbf{e}_{z}$,
respectively. This can be simplified, if $\Gamma _{2}=\Gamma _{1}\equiv
\Gamma $, then Eq.~(\ref{Bloch_vector}) becomes Eq.~(\ref{eq4x}).

In the case of \textit{free evolution}, when $A=0$, with long relaxation
times, the Bloch equation can be written for the diagonal and off-diagonal
density matrix components, respectively $\rho _{00}=\frac{1}{2}\left(
1+Z\right) $ and $\rho _{10}=\frac{1}{2}\left( X+iY\right) $:%
\begin{eqnarray}
\dot{\rho}_{10} &=&-i\omega _{\mathrm{q}}\rho _{10}, \\
\dot{\rho}_{00} &=&0.  \notag
\end{eqnarray}%
The solution is described by the constant energy-level populations (defined
by the initial condition) and the beating, with frequency $\omega _{\mathrm{q%
}}$, of the off-diagonal components:%
\begin{eqnarray}
\rho _{00}(t) &=&\rho _{00}(0)=\mathrm{const}, \\
\rho _{10}(t) &=&\rho _{10}(0)\exp \left( -i\omega _{\mathrm{q}}t\right) .
\notag
\end{eqnarray}

\begin{figure*}[th]
\includegraphics[width=16cm]{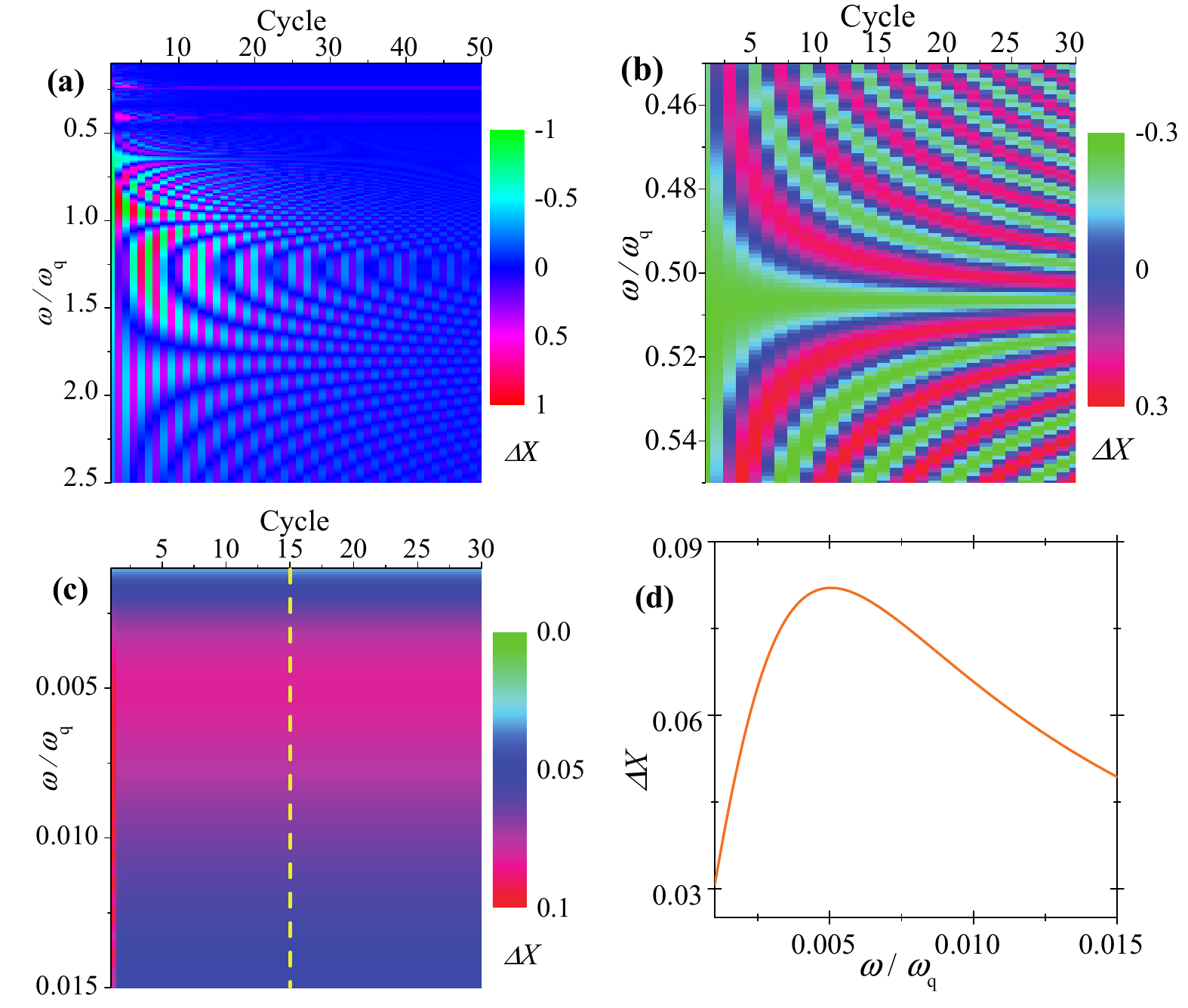}
\caption{(a)~Hysteresis of $X$ as a function of $\protect%
\omega $ and cycle calculated using Eq.~(\protect\ref{DX}). This plot was
obtained employing the following set of parameters: $\protect\varepsilon %
_{0}=0$, $\Omega _{\mathrm{R}}^{(0)}/\protect\omega _{\mathrm{q}}=1/2$, $%
\Gamma =0.01\Delta /\hbar $. (b)~Two-photon excitation regime. The parameter
values are similar to the ones in Fig.~\protect\ref{Fig:two-photon}.
(c)~Hysteresis of $X$ in the delayed-response region found at $\protect%
\varepsilon _{0}=0$, $\Omega _{\mathrm{R}}^{(0)}/\protect\omega _{\mathrm{q}%
}=1/2$, and $\Gamma =0.01\Delta /\hbar $. (d)~$\Delta X$ along the vertical
cross-section of (c) denoted by the dashed line.}
\label{Fig:map5}
\end{figure*}

Consider now another situation, when the transition between the qubit energy
levels is induced by means of the Rabi oscillations under\textbf{\ }\textit{%
resonant driving}, when $\delta \omega \equiv \omega -\omega _{\mathrm{q}%
}\ll \omega $. Then we make the transformation
\begin{equation}
\widetilde{\rho }_{10}=\rho _{10}\exp \left( i\omega t\right) \equiv
\widetilde{X}+i\widetilde{Y}  \label{rho_tilde}
\end{equation}%
(the diagonal component $Z$ is left unchanged) and use the rotating-wave
approximation (RWA), i.e. neglect the fast-rotating terms, and from (\ref%
{Bloch}) we obtain%
\begin{eqnarray}
\dot{Z} &=&-\Omega _{\mathrm{R}}^{(0)}\widetilde{X}-\Gamma _{1}(Z-Z_{0}),
\label{i} \\
\overset{.}{\widetilde{\rho }}_{10} &=&\left( i\delta \omega -\Gamma
_{2}\right) \widetilde{\rho }_{10}+\frac{1}{2}\Omega _{\mathrm{R}}^{(0)}Z.
\end{eqnarray}%
The latter equation can be rewritten as:%
\begin{eqnarray}
\overset{\cdot }{\widetilde{X}} &=&-\delta \omega \widetilde{Y}+\Omega _{%
\mathrm{R}}^{(0)}Z-\Gamma _{2}\widetilde{X},  \label{ii} \\
\overset{\cdot }{\widetilde{Y}} &=&\delta \omega \widetilde{X}-\Gamma _{2}%
\widetilde{Y}.  \label{iii}
\end{eqnarray}%
The system of equations (\ref{i}, \ref{ii}, \ref{iii}) can be solved
analytically in two cases: for the stationary case, at times $t\gg
T_{1},T_{2}$ and for the case of long relaxation rates, $T_{1},T_{2}%
\longrightarrow \infty $. The former solution is obtained by simply assuming
the l.h.s.~of those equations being zero. Consider now in more detail the
latter solution to see how Rabi oscillations emerge. In this case we ignore
the relaxation terms, and then look for the partial solutions of the
differential equations. Let us write down the solutions here for the initial
condition of the qubit being in the ground state with $\widetilde{X}(0)=%
\widetilde{Y}(0)=0$ and $Z(0)=1$:%
\begin{eqnarray}
\widetilde{X}(t) &=&\frac{\Omega _{\mathrm{R}}^{(0)}}{\Omega _{\mathrm{R}}}%
\sin \Omega _{\mathrm{R}}t,  \label{x} \\
\widetilde{Y}(t) &=&\frac{\delta \omega \Omega _{\mathrm{R}}^{(0)}}{\Omega _{%
\mathrm{R}}^{2}}\left( 1-\cos \Omega _{\mathrm{R}}t\right) ,  \label{y} \\
Z(t) &=&1-\frac{\Omega _{\mathrm{R}}^{(0)2}}{\Omega _{\mathrm{R}}^{2}}\left(
1-\cos \Omega _{\mathrm{R}}t\right) ,  \label{z} \\
\Omega _{\mathrm{R}} &=&\sqrt{\Omega _{\mathrm{R}}^{(0)2}+\delta \omega ^{2}}%
.
\end{eqnarray}%
This provides the formula for the Rabi oscillations (of the upper-level
occupation probability):%
\begin{eqnarray}
P_{+}(t) &=&\frac{1}{2}(1-Z)=\overline{P_{+}}\left( 1-\cos \Omega _{\mathrm{R%
}}t\right) ,\text{ \ }  \label{Rabi} \\
\overline{P_{+}} &=&\frac{\Omega _{\mathrm{R}}^{(0)2}}{2\Omega _{\mathrm{R}%
}^{2}}=\frac{1}{2}\frac{\Omega _{\mathrm{R}}^{(0)2}}{\Omega _{\mathrm{R}%
}^{(0)2}+\delta \omega ^{2}}.
\end{eqnarray}

For the qubit current in Eq.~(\ref{Iq}), we need $Z$ and $X$. The former
value is given by Eq.~(\ref{z}) and the latter is found with Eq.~(\ref%
{rho_tilde}):%
\begin{equation}
X=\frac{\Omega _{\mathrm{R}}^{(0)}}{\Omega _{\mathrm{R}}}\left( \sin \Omega
_{\mathrm{R}}t\cos \omega t+\frac{\delta \omega }{\Omega _{\mathrm{R}}}%
\left( 1-\cos \Omega _{\mathrm{R}}t\right) \sin \omega t\right) .
\end{equation}%
In particular, in resonance, at $\delta \omega =0$ (then $\Omega _{\mathrm{R}%
}=\Omega _{\mathrm{R}}^{(0)}$), we obtain:%
\begin{equation}
\frac{I_{\mathrm{q}}}{I_{\mathrm{p}}}=\frac{\Delta }{\Delta E}\sin \Omega _{%
\mathrm{R}}^{(0)}t\cos \omega t-\frac{\varepsilon _{0}}{\Delta E}\cos \Omega
_{\mathrm{R}}^{(0)}t\text{.}
\end{equation}%
This is further discussed in the main text, Sec.~IV.

Thus we have analyzed the Rabi oscillations. It is worth pointing out that
these oscillations can be viewed as a consequence of the constructive
interference of the Landau-Zener-St\"{u}ckelberg-Majorana (LZSM)
transitions~[\onlinecite{Shevchenko10, Zhou14}]. The opposite case refers to
the destructive LZSM interference, which corresponds to the periodic small
rising, in the relevant adiabatic basis, of the height given by the LZSM
probability. Also we note that besides the sinusoidal driving considered
here in detail, there are also other aspects of the qubits driven by
different pulses~[\onlinecite{Wendin07}], which may also be important in the
context of quantum memory devices.

\section{Hysteresis size}

\label{SMplots}

In addition to the results presented in Sec.~\ref{sec4}, here we provide
some additional plots (Fig.~\ref{Fig:map5}) that could help to better
understand the system's response (in certain ranges of parameters). These
plots were obtained similarly to the plots in Fig.~\ref{Fig:map1}.

In particular, Fig.~\ref{Fig:map5}(a) exemplifies the Rabi oscillations
regime (Sec.~\ref{sec:Rabi}) for the case of $\Omega _{\mathrm{R}%
}^{(0)}/\omega _{\mathrm{q}}=1/2$. As we previously discussed, for a certain
$\omega \sim \omega _{\mathrm{q}}$, the period of the hysteresis loop is
double the period of the external excitation [see also Fig.~\ref{Fig:Rabi}%
(a)]. This feature is clearly seen in Fig.~\ref{Fig:map5}(a) as an
alternation of the hysteresis sign in the consecutive cycles.

Figure~\ref{Fig:map5}(b) is related to the two-photon excitation regime
(Sec.~\ref{2ph_exc}) showing that a stable hysteresis indeed occurs at a
certain $\omega \sim \omega _{\mathrm{q}}/2$. The oscillations in the
hysteresis appear when $\omega $ moves up or down from the hysteretic value.
In Fig.~\ref{Fig:map5}(c) we additionally explore the delayed-response
mechanism of the hysteresis (Sec.~\ref{del_resp}). According to Fig.~\ref%
{Fig:map5}(c), the delayed-response mechanism provides a stable hysteresis
with a fixed sign. Its maximum corresponds to $\omega \sim \Gamma /2$ as
shown in Fig. \ref{Fig:map5}(d).

\section{Quantum uncertainty}

Let us finally calculate the quantum uncertainty \cite{Landau1981QM} of the
measurement of output. For this purpose, we introduce the output operator $%
\hat{y}(t)$ as (see Eq. (\ref{u(t)}))
\begin{equation}
\hat{y}(t)=y_{0}\frac{u(t)}{u_{0}}-y_{0}\varkappa \sigma _{z}.
\label{yhat(t)}
\end{equation}%
Then, the standard deviation of $y$ is given by
\begin{equation}
\Delta y(t)=\sqrt{\langle \left( \hat{y}(t)-y(t)\right) ^{2}\rangle }=\sqrt{%
\langle \hat{y}^{2}(t)\rangle -y^{2}(t)},  \label{sigma(t)}
\end{equation}%
where $y(t)=\langle \hat{y}(t)\rangle $. Using Eq. (\ref{yhat(t)}), we
finally obtain
\begin{equation}
\Delta y(t)=y_{0}\varkappa \sqrt{1-\langle \sigma _{z}\rangle ^{2}}.
\label{sigma1(t)}
\end{equation}

\begin{figure}[th]
\includegraphics[width=8.7cm]{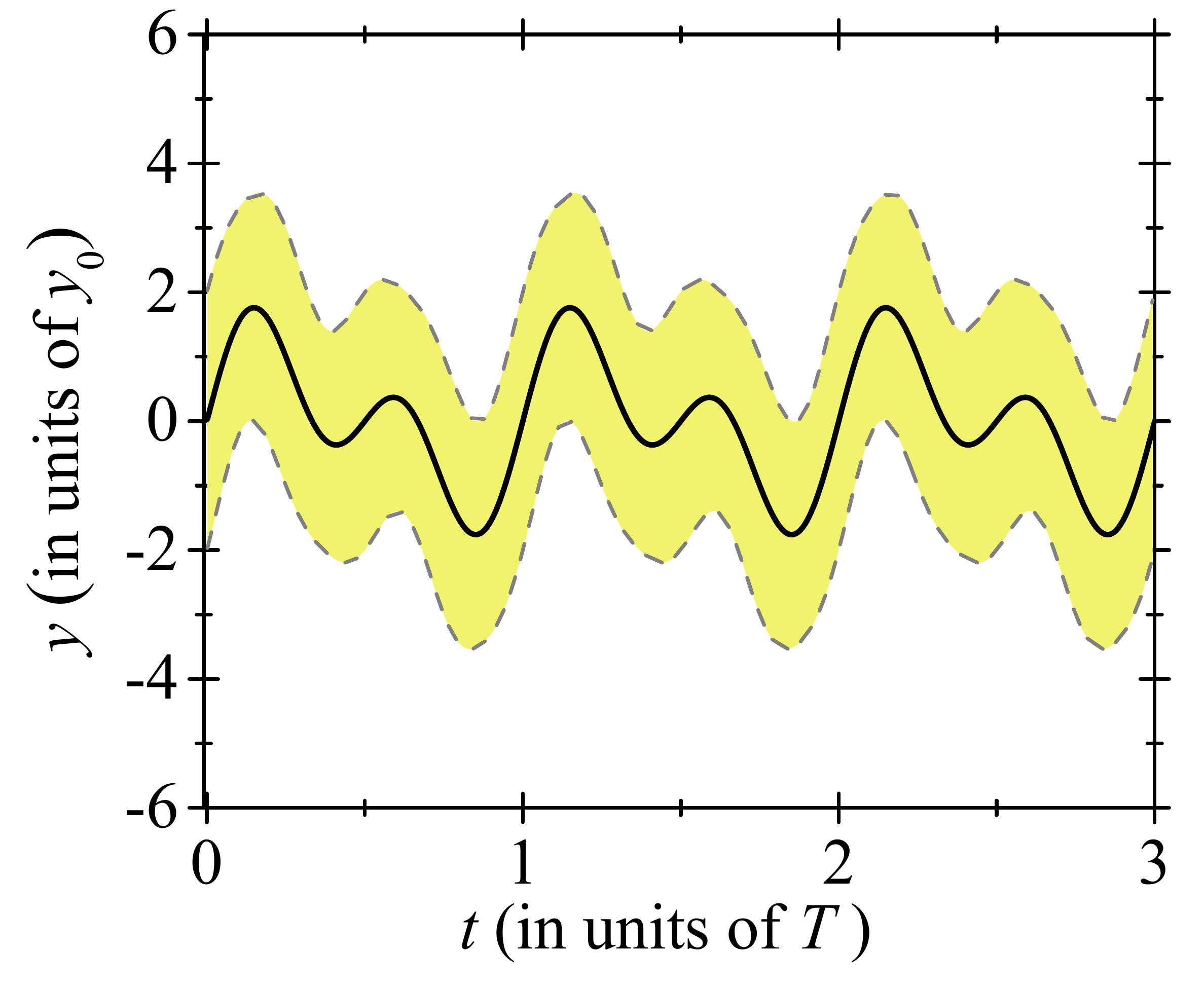}
\caption{Time-dependence of the output $y$ and its
uncertainty region (the dashed lines represent $y(t)\pm \Delta y(t)$).}
\label{Fig:Rabi1}
\end{figure}

Let us illustrate Eq. (\ref{sigma1(t)}). For this purpose, we consider the
Rabi oscillations example from Sec. \ref{sec:Rabi}. Assuming the case $%
\varepsilon_0=0$ (Eq. (\ref{sigma_Z1})) and $\Omega^R/\omega=1$ (Fig. \ref%
{Fig:Rabi}), we get
\begin{equation}
\Delta y(t)=y_0\varkappa\sqrt{1-\sin^2 \Omega_{\mathrm{R}}^{(0)}t\cos^2
\omega t}
\end{equation}
that is illustrated in Fig. \ref{Fig:Rabi1}.

\nocite{apsrev41Control}
\bibliographystyle{apsrev4-1}
\bibliography{memcapacitor}

\begin{thebibliography}{37}%
\makeatletter
\providecommand \@ifxundefined [1]{%
 \@ifx{#1\undefined}
}%
\providecommand \@ifnum [1]{%
 \ifnum #1\expandafter \@firstoftwo
 \else \expandafter \@secondoftwo
 \fi
}%
\providecommand \@ifx [1]{%
 \ifx #1\expandafter \@firstoftwo
 \else \expandafter \@secondoftwo
 \fi
}%
\providecommand \natexlab [1]{#1}%
\providecommand \enquote  [1]{``#1''}%
\providecommand \bibnamefont  [1]{#1}%
\providecommand \bibfnamefont [1]{#1}%
\providecommand \citenamefont [1]{#1}%
\providecommand \href@noop [0]{\@secondoftwo}%
\providecommand \href [0]{\begingroup \@sanitize@url \@href}%
\providecommand \@href[1]{\@@startlink{#1}\@@href}%
\providecommand \@@href[1]{\endgroup#1\@@endlink}%
\providecommand \@sanitize@url [0]{\catcode `\\12\catcode `\$12\catcode
  `\&12\catcode `\#12\catcode `\^12\catcode `\_12\catcode `\%12\relax}%
\providecommand \@@startlink[1]{}%
\providecommand \@@endlink[0]{}%
\providecommand \url  [0]{\begingroup\@sanitize@url \@url }%
\providecommand \@url [1]{\endgroup\@href {#1}{\urlprefix }}%
\providecommand \urlprefix  [0]{URL }%
\providecommand \Eprint [0]{\href }%
\providecommand \doibase [0]{http://dx.doi.org/}%
\providecommand \selectlanguage [0]{\@gobble}%
\providecommand \bibinfo  [0]{\@secondoftwo}%
\providecommand \bibfield  [0]{\@secondoftwo}%
\providecommand \translation [1]{[#1]}%
\providecommand \BibitemOpen [0]{}%
\providecommand \bibitemStop [0]{}%
\providecommand \bibitemNoStop [0]{.\EOS\space}%
\providecommand \EOS [0]{\spacefactor3000\relax}%
\providecommand \BibitemShut  [1]{\csname bibitem#1\endcsname}%
\let\auto@bib@innerbib\@empty
%</preamble>
\bibitem [{\citenamefont {Chua}(1971)}]{chua71a}%
  \BibitemOpen
  \bibfield  {author} {\bibinfo {author} {\bibfnamefont {L.~O.}\ \bibnamefont
  {Chua}},\ }\bibfield  {title} {\enquote {\bibinfo {title} {Memristor - the
  missing circuit element},}\ }\href@noop {} {\bibfield  {journal} {\bibinfo
  {journal} {{IEEE} Trans. Circuit Theory}\ }\textbf {\bibinfo {volume} {18}},\
  \bibinfo {pages} {507} (\bibinfo {year} {1971})}\BibitemShut {NoStop}%
\bibitem [{\citenamefont {Chua}\ and\ \citenamefont {Kang}(1976)}]{chua76a}%
  \BibitemOpen
  \bibfield  {author} {\bibinfo {author} {\bibfnamefont {L.~O.}\ \bibnamefont
  {Chua}}\ and\ \bibinfo {author} {\bibfnamefont {S.~M.}\ \bibnamefont
  {Kang}},\ }\bibfield  {title} {\enquote {\bibinfo {title} {Memristive devices
  and systems},}\ }\href@noop {} {\bibfield  {journal} {\bibinfo  {journal}
  {Proc. {IEEE}}\ }\textbf {\bibinfo {volume} {64}},\ \bibinfo {pages} {209}
  (\bibinfo {year} {1976})}\BibitemShut {NoStop}%
\bibitem [{\citenamefont {{Di Ventra}}\ \emph {et~al.}(2009)\citenamefont {{Di
  Ventra}}, \citenamefont {Pershin},\ and\ \citenamefont {Chua}}]{diventra09a}%
  \BibitemOpen
  \bibfield  {author} {\bibinfo {author} {\bibfnamefont {M.}~\bibnamefont {{Di
  Ventra}}}, \bibinfo {author} {\bibfnamefont {Y.~V.}\ \bibnamefont {Pershin}},
  \ and\ \bibinfo {author} {\bibfnamefont {L.~O.}\ \bibnamefont {Chua}},\
  }\bibfield  {title} {\enquote {\bibinfo {title} {Circuit elements with
  memory: Memristors, memcapacitors, and meminductors},}\ }\href@noop {}
  {\bibfield  {journal} {\bibinfo  {journal} {Proc. {IEEE}}\ }\textbf {\bibinfo
  {volume} {97}},\ \bibinfo {pages} {1717} (\bibinfo {year}
  {2009})}\BibitemShut {NoStop}%
\bibitem [{\citenamefont {Pershin}\ and\ \citenamefont
  {Di~Ventra}(2011)}]{pershin11a}%
  \BibitemOpen
  \bibfield  {author} {\bibinfo {author} {\bibfnamefont {Y.~V.}\ \bibnamefont
  {Pershin}}\ and\ \bibinfo {author} {\bibfnamefont {M.}~\bibnamefont
  {Di~Ventra}},\ }\bibfield  {title} {\enquote {\bibinfo {title} {Memory
  effects in complex materials and nanoscale systems},}\ }\href@noop {}
  {\bibfield  {journal} {\bibinfo  {journal} {Advances in Physics}\ }\textbf
  {\bibinfo {volume} {60}},\ \bibinfo {pages} {145} (\bibinfo {year}
  {2011})}\BibitemShut {NoStop}%
\bibitem [{\citenamefont {Cohen}\ \emph {et~al.}(2012)\citenamefont {Cohen},
  \citenamefont {Pershin},\ and\ \citenamefont {Di~Ventra}}]{Cohen12a}%
  \BibitemOpen
  \bibfield  {author} {\bibinfo {author} {\bibfnamefont {G.~Z.}\ \bibnamefont
  {Cohen}}, \bibinfo {author} {\bibfnamefont {Y.~V.}\ \bibnamefont {Pershin}},
  \ and\ \bibinfo {author} {\bibfnamefont {M.}~\bibnamefont {Di~Ventra}},\
  }\bibfield  {title} {\enquote {\bibinfo {title} {Lagrange formalism of memory
  circuit elements: {C}lassical and quantum formulations},}\ }\href@noop {}
  {\bibfield  {journal} {\bibinfo  {journal} {Phys. Rev. B}\ }\textbf {\bibinfo
  {volume} {85}},\ \bibinfo {pages} {165428} (\bibinfo {year}
  {2012})}\BibitemShut {NoStop}%
\bibitem [{\citenamefont {Gough}\ and\ \citenamefont {Zhang}(2015)}]{Gough15}%
  \BibitemOpen
  \bibfield  {author} {\bibinfo {author} {\bibfnamefont {J.~E.}\ \bibnamefont
  {Gough}}\ and\ \bibinfo {author} {\bibfnamefont {G.}~\bibnamefont {Zhang}},\
  }\bibfield  {title} {\enquote {\bibinfo {title} {Classical and quantum
  stochastic models of resistive and memristive circuits},}\ }\href@noop {}
  {\bibfield  {journal} {\bibinfo  {journal} {arXiv:1510.08243}\ } (\bibinfo
  {year} {2015})}\BibitemShut {NoStop}%
\bibitem [{\citenamefont {Pfeiffer}\ \emph {et~al.}(2015)\citenamefont
  {Pfeiffer}, \citenamefont {Egusquiza}, \citenamefont {Di~Ventra},
  \citenamefont {Sanz},\ and\ \citenamefont {Solano}}]{Pfeiffer15a}%
  \BibitemOpen
  \bibfield  {author} {\bibinfo {author} {\bibfnamefont {P.}~\bibnamefont
  {Pfeiffer}}, \bibinfo {author} {\bibfnamefont {I.~L.}\ \bibnamefont
  {Egusquiza}}, \bibinfo {author} {\bibfnamefont {M.}~\bibnamefont
  {Di~Ventra}}, \bibinfo {author} {\bibfnamefont {M.}~\bibnamefont {Sanz}}, \
  and\ \bibinfo {author} {\bibfnamefont {E.}~\bibnamefont {Solano}},\
  }\bibfield  {title} {\enquote {\bibinfo {title} {Quantum memristors},}\
  }\href@noop {} {\bibfield  {journal} {\bibinfo  {journal} {arXiv:1511.02192}\
  } (\bibinfo {year} {2015})}\BibitemShut {NoStop}%
\bibitem [{\citenamefont {Likharev}(1986)}]{Likharev}%
  \BibitemOpen
  \bibfield  {author} {\bibinfo {author} {\bibfnamefont {K.~K.}\ \bibnamefont
  {Likharev}},\ }\href@noop {} {\emph {\bibinfo {title} {Dynamics of
  {J}osephson Junctions and Circuits}}}\ (\bibinfo  {publisher} {Gordon and
  Breach, New York},\ \bibinfo {year} {1986})\BibitemShut {NoStop}%
\bibitem [{\citenamefont {Wendin}\ and\ \citenamefont
  {Shumeiko}(2007)}]{Wendin07}%
  \BibitemOpen
  \bibfield  {author} {\bibinfo {author} {\bibfnamefont {G.}~\bibnamefont
  {Wendin}}\ and\ \bibinfo {author} {\bibfnamefont {V.~S.}\ \bibnamefont
  {Shumeiko}},\ }\bibfield  {title} {\enquote {\bibinfo {title} {Quantum bits
  with {J}osephson junctions},}\ }\href@noop {} {\bibfield  {journal} {\bibinfo
   {journal} {Low Temp. Phys.}\ }\textbf {\bibinfo {volume} {33}},\ \bibinfo
  {pages} {724} (\bibinfo {year} {2007})}\BibitemShut {NoStop}%
\bibitem [{\citenamefont {You}\ and\ \citenamefont {Nori}(2011)}]{You11}%
  \BibitemOpen
  \bibfield  {author} {\bibinfo {author} {\bibfnamefont {J.~Q.}\ \bibnamefont
  {You}}\ and\ \bibinfo {author} {\bibfnamefont {F.}~\bibnamefont {Nori}},\
  }\bibfield  {title} {\enquote {\bibinfo {title} {Atomic physics and quantum
  optics using superconducting circuits},}\ }\href@noop {} {\bibfield
  {journal} {\bibinfo  {journal} {Nature}\ }\textbf {\bibinfo {volume} {474}},\
  \bibinfo {pages} {589} (\bibinfo {year} {2011})}\BibitemShut {NoStop}%
\bibitem [{\citenamefont {Xiang}\ \emph {et~al.}(2013)\citenamefont {Xiang},
  \citenamefont {Ashhab}, \citenamefont {You},\ and\ \citenamefont
  {Nori}}]{Xiang13}%
  \BibitemOpen
  \bibfield  {author} {\bibinfo {author} {\bibfnamefont {Z.-L.}\ \bibnamefont
  {Xiang}}, \bibinfo {author} {\bibfnamefont {S.}~\bibnamefont {Ashhab}},
  \bibinfo {author} {\bibfnamefont {J.~Q.}\ \bibnamefont {You}}, \ and\
  \bibinfo {author} {\bibfnamefont {F.}~\bibnamefont {Nori}},\ }\bibfield
  {title} {\enquote {\bibinfo {title} {Hybrid quantum circuits:
  {S}uperconducting circuits interacting with other quantum systems},}\
  }\href@noop {} {\bibfield  {journal} {\bibinfo  {journal} {Rev. Mod. Phys.}\
  }\textbf {\bibinfo {volume} {85}},\ \bibinfo {pages} {623} (\bibinfo {year}
  {2013})}\BibitemShut {NoStop}%
\bibitem [{\citenamefont {Chua}(2003)}]{chua03a}%
  \BibitemOpen
  \bibfield  {author} {\bibinfo {author} {\bibfnamefont {L.~O.}\ \bibnamefont
  {Chua}},\ }\bibfield  {title} {\enquote {\bibinfo {title} {Nonlinear circuit
  foundations for nanodevices, part i: The four-element torus},}\ }\href@noop
  {} {\bibfield  {journal} {\bibinfo  {journal} {Proc. {IEEE}}\ }\textbf
  {\bibinfo {volume} {91}},\ \bibinfo {pages} {1830} (\bibinfo {year}
  {2003})}\BibitemShut {NoStop}%
\bibitem [{\citenamefont {Peotta}\ and\ \citenamefont
  {Di~Ventra}(2014)}]{Peotta14}%
  \BibitemOpen
  \bibfield  {author} {\bibinfo {author} {\bibfnamefont {S.}~\bibnamefont
  {Peotta}}\ and\ \bibinfo {author} {\bibfnamefont {M.}~\bibnamefont
  {Di~Ventra}},\ }\bibfield  {title} {\enquote {\bibinfo {title}
  {Superconducting memristors},}\ }\href@noop {} {\bibfield  {journal}
  {\bibinfo  {journal} {Phys. Rev. Applied}\ }\textbf {\bibinfo {volume} {2}},\
  \bibinfo {pages} {034011} (\bibinfo {year} {2014})}\BibitemShut {NoStop}%
\bibitem [{\citenamefont {Barone}\ and\ \citenamefont
  {Paterno}(1982)}]{Barone}%
  \BibitemOpen
  \bibfield  {author} {\bibinfo {author} {\bibfnamefont {A.}~\bibnamefont
  {Barone}}\ and\ \bibinfo {author} {\bibfnamefont {G.}~\bibnamefont
  {Paterno}},\ }\href@noop {} {\emph {\bibinfo {title} {Physics and Application
  of the {J}osephson Effect}}}\ (\bibinfo  {publisher} {Wiley, New York},\
  \bibinfo {year} {1982})\BibitemShut {NoStop}%
\bibitem [{\citenamefont {Martinez-Rincon}\ \emph {et~al.}(2010)\citenamefont
  {Martinez-Rincon}, \citenamefont {Di~Ventra},\ and\ \citenamefont
  {Pershin}}]{martinez09a}%
  \BibitemOpen
  \bibfield  {author} {\bibinfo {author} {\bibfnamefont {J.}~\bibnamefont
  {Martinez-Rincon}}, \bibinfo {author} {\bibfnamefont {Massimiliano}\
  \bibnamefont {Di~Ventra}}, \ and\ \bibinfo {author} {\bibfnamefont
  {Yuriy~V.}\ \bibnamefont {Pershin}},\ }\bibfield  {title} {\enquote {\bibinfo
  {title} {Solid-state memcapacitive system with negative and diverging
  capacitance},}\ }\href@noop {} {\bibfield  {journal} {\bibinfo  {journal}
  {Phys. Rev. B}\ }\textbf {\bibinfo {volume} {81}},\ \bibinfo {pages} {195430}
  (\bibinfo {year} {2010})}\BibitemShut {NoStop}%
\bibitem [{\citenamefont {Shevchenko}\ \emph {et~al.}(2010)\citenamefont
  {Shevchenko}, \citenamefont {Ashhab},\ and\ \citenamefont
  {Nori}}]{Shevchenko10}%
  \BibitemOpen
  \bibfield  {author} {\bibinfo {author} {\bibfnamefont {S.N.}\ \bibnamefont
  {Shevchenko}}, \bibinfo {author} {\bibfnamefont {S.}~\bibnamefont {Ashhab}},
  \ and\ \bibinfo {author} {\bibfnamefont {F.}~\bibnamefont {Nori}},\
  }\bibfield  {title} {\enquote {\bibinfo {title}
  {Landau-{Z}ener-{S}t{\"u}ckelberg interferometry},}\ }\href@noop {}
  {\bibfield  {journal} {\bibinfo  {journal} {Phys. Rep.}\ }\textbf {\bibinfo
  {volume} {492}},\ \bibinfo {pages} {1} (\bibinfo {year} {2010})}\BibitemShut
  {NoStop}%
\bibitem [{\citenamefont {Salmilehto}\ \emph {et~al.}(2016)\citenamefont
  {Salmilehto}, \citenamefont {Deppe}, \citenamefont {Di~Ventra}, \citenamefont
  {Sanz},\ and\ \citenamefont {Solano}}]{Salmilehto16}%
  \BibitemOpen
  \bibfield  {author} {\bibinfo {author} {\bibfnamefont {J.}~\bibnamefont
  {Salmilehto}}, \bibinfo {author} {\bibfnamefont {F.}~\bibnamefont {Deppe}},
  \bibinfo {author} {\bibfnamefont {M.}~\bibnamefont {Di~Ventra}}, \bibinfo
  {author} {\bibfnamefont {M.}~\bibnamefont {Sanz}}, \ and\ \bibinfo {author}
  {\bibfnamefont {E.}~\bibnamefont {Solano}},\ }\bibfield  {title} {\enquote
  {\bibinfo {title} {Quantum memristors with superconducting circuits},}\
  }\href@noop {} {\bibfield  {journal} {\bibinfo  {journal} {arXiv:1603.04487}\
  } (\bibinfo {year} {2016})}\BibitemShut {NoStop}%
\bibitem [{\citenamefont {Shevchenko}\ \emph {et~al.}(2014)\citenamefont
  {Shevchenko}, \citenamefont {Oelsner}, \citenamefont {Greenberg},
  \citenamefont {Macha}, \citenamefont {Karpov}, \citenamefont {Grajcar},
  \citenamefont {H\"ubner}, \citenamefont {Omelyanchouk},\ and\ \citenamefont
  {Il'ichev}}]{Shevchenko14}%
  \BibitemOpen
  \bibfield  {author} {\bibinfo {author} {\bibfnamefont {S.~N.}\ \bibnamefont
  {Shevchenko}}, \bibinfo {author} {\bibfnamefont {G.}~\bibnamefont {Oelsner}},
  \bibinfo {author} {\bibfnamefont {Ya.~S.}\ \bibnamefont {Greenberg}},
  \bibinfo {author} {\bibfnamefont {P.}~\bibnamefont {Macha}}, \bibinfo
  {author} {\bibfnamefont {D.~S.}\ \bibnamefont {Karpov}}, \bibinfo {author}
  {\bibfnamefont {M.}~\bibnamefont {Grajcar}}, \bibinfo {author} {\bibfnamefont
  {U.}~\bibnamefont {H\"ubner}}, \bibinfo {author} {\bibfnamefont {A.~N.}\
  \bibnamefont {Omelyanchouk}}, \ and\ \bibinfo {author} {\bibfnamefont
  {E.}~\bibnamefont {Il'ichev}},\ }\bibfield  {title} {\enquote {\bibinfo
  {title} {Amplification and attenuation of a probe signal by doubly dressed
  states},}\ }\href@noop {} {\bibfield  {journal} {\bibinfo  {journal} {Phys.
  Rev. B}\ }\textbf {\bibinfo {volume} {89}},\ \bibinfo {pages} {184504}
  (\bibinfo {year} {2014})}\BibitemShut {NoStop}%
\bibitem [{\citenamefont {Sillanp\"a\"a}\ \emph {et~al.}(2005)\citenamefont
  {Sillanp\"a\"a}, \citenamefont {Lehtinen}, \citenamefont {Paila},
  \citenamefont {Makhlin}, \citenamefont {Roschier},\ and\ \citenamefont
  {Hakonen}}]{Sillanpaa05}%
  \BibitemOpen
  \bibfield  {author} {\bibinfo {author} {\bibfnamefont {M.~A.}\ \bibnamefont
  {Sillanp\"a\"a}}, \bibinfo {author} {\bibfnamefont {T.}~\bibnamefont
  {Lehtinen}}, \bibinfo {author} {\bibfnamefont {A.}~\bibnamefont {Paila}},
  \bibinfo {author} {\bibfnamefont {Y.}~\bibnamefont {Makhlin}}, \bibinfo
  {author} {\bibfnamefont {L.}~\bibnamefont {Roschier}}, \ and\ \bibinfo
  {author} {\bibfnamefont {P.~J.}\ \bibnamefont {Hakonen}},\ }\bibfield
  {title} {\enquote {\bibinfo {title} {Direct observation of {J}osephson
  capacitance},}\ }\href@noop {} {\bibfield  {journal} {\bibinfo  {journal}
  {Phys. Rev. Lett.}\ }\textbf {\bibinfo {volume} {95}},\ \bibinfo {pages}
  {206806} (\bibinfo {year} {2005})}\BibitemShut {NoStop}%
\bibitem [{\citenamefont {Duty}\ \emph {et~al.}(2005)\citenamefont {Duty},
  \citenamefont {Johansson}, \citenamefont {Bladh}, \citenamefont {Gunnarsson},
  \citenamefont {Wilson},\ and\ \citenamefont {Delsing}}]{Duty05}%
  \BibitemOpen
  \bibfield  {author} {\bibinfo {author} {\bibfnamefont {T.}~\bibnamefont
  {Duty}}, \bibinfo {author} {\bibfnamefont {G.}~\bibnamefont {Johansson}},
  \bibinfo {author} {\bibfnamefont {K.}~\bibnamefont {Bladh}}, \bibinfo
  {author} {\bibfnamefont {D.}~\bibnamefont {Gunnarsson}}, \bibinfo {author}
  {\bibfnamefont {C.}~\bibnamefont {Wilson}}, \ and\ \bibinfo {author}
  {\bibfnamefont {P.}~\bibnamefont {Delsing}},\ }\bibfield  {title} {\enquote
  {\bibinfo {title} {Observation of quantum capacitance in the {C}ooper-pair
  transistor},}\ }\href@noop {} {\bibfield  {journal} {\bibinfo  {journal}
  {Phys. Rev. Lett.}\ }\textbf {\bibinfo {volume} {95}},\ \bibinfo {pages}
  {206807} (\bibinfo {year} {2005})}\BibitemShut {NoStop}%
\bibitem [{\citenamefont {Johansson}\ \emph {et~al.}(2006)\citenamefont
  {Johansson}, \citenamefont {Tornberg}, \citenamefont {Shumeiko},\ and\
  \citenamefont {Wendin}}]{Johansson06}%
  \BibitemOpen
  \bibfield  {author} {\bibinfo {author} {\bibfnamefont {G.}~\bibnamefont
  {Johansson}}, \bibinfo {author} {\bibfnamefont {L.}~\bibnamefont {Tornberg}},
  \bibinfo {author} {\bibfnamefont {V.~S.}\ \bibnamefont {Shumeiko}}, \ and\
  \bibinfo {author} {\bibfnamefont {G.}~\bibnamefont {Wendin}},\ }\bibfield
  {title} {\enquote {\bibinfo {title} {Readout methods and devices for
  {J}osephson-junction-based solid-state qubits},}\ }\href@noop {} {\bibfield
  {journal} {\bibinfo  {journal} {J. Phys. Cond. Matt.}\ }\textbf {\bibinfo
  {volume} {18}},\ \bibinfo {pages} {S901} (\bibinfo {year}
  {2006})}\BibitemShut {NoStop}%
\bibitem [{\citenamefont {Makhlin}\ \emph {et~al.}(2002)\citenamefont
  {Makhlin}, \citenamefont {Sch{\"o}n},\ and\ \citenamefont
  {Shnirman}}]{Makhlin02}%
  \BibitemOpen
  \bibfield  {author} {\bibinfo {author} {\bibfnamefont {Y.}~\bibnamefont
  {Makhlin}}, \bibinfo {author} {\bibfnamefont {G.}~\bibnamefont {Sch{\"o}n}},
  \ and\ \bibinfo {author} {\bibfnamefont {A.}~\bibnamefont {Shnirman}},\
  }\bibfield  {title} {\enquote {\bibinfo {title} {Josephson quantum bits in
  the flux regime},}\ }\href@noop {} {\bibfield  {journal} {\bibinfo  {journal}
  {Physica C}\ }\textbf {\bibinfo {volume} {368}},\ \bibinfo {pages} {276}
  (\bibinfo {year} {2002})}\BibitemShut {NoStop}%
\bibitem [{\citenamefont {Blum}(1981)}]{Blum}%
  \BibitemOpen
  \bibfield  {author} {\bibinfo {author} {\bibfnamefont {K.}~\bibnamefont
  {Blum}},\ }\href@noop {} {\emph {\bibinfo {title} {Density Matrix Theory and
  Applications}}}\ (\bibinfo  {publisher} {Plenum, New York},\ \bibinfo {year}
  {1981})\BibitemShut {NoStop}%
\bibitem [{\citenamefont {Bloch}\ and\ \citenamefont
  {Siegert}(1940)}]{Bloch40a}%
  \BibitemOpen
  \bibfield  {author} {\bibinfo {author} {\bibfnamefont {F.}~\bibnamefont
  {Bloch}}\ and\ \bibinfo {author} {\bibfnamefont {A.}~\bibnamefont
  {Siegert}},\ }\bibfield  {title} {\enquote {\bibinfo {title} {Magnetic
  resonance for nonrotating fields},}\ }\href@noop {} {\bibfield  {journal}
  {\bibinfo  {journal} {Phys. Rev.}\ }\textbf {\bibinfo {volume} {57}},\
  \bibinfo {pages} {522} (\bibinfo {year} {1940})}\BibitemShut {NoStop}%
\bibitem [{\citenamefont {Shevchenko}\ \emph {et~al.}(2008)\citenamefont
  {Shevchenko}, \citenamefont {Omelyanchouk}, \citenamefont {Zagoskin},
  \citenamefont {Il'ichev},\ and\ \citenamefont {Nori}}]{Shevchenko08}%
  \BibitemOpen
  \bibfield  {author} {\bibinfo {author} {\bibfnamefont {S.~N.}\ \bibnamefont
  {Shevchenko}}, \bibinfo {author} {\bibfnamefont {A.~N.}\ \bibnamefont
  {Omelyanchouk}}, \bibinfo {author} {\bibfnamefont {A.~M.}\ \bibnamefont
  {Zagoskin}}, \bibinfo {author} {\bibfnamefont {E.}~\bibnamefont {Il'ichev}},
  \ and\ \bibinfo {author} {\bibfnamefont {F.}~\bibnamefont {Nori}},\
  }\bibfield  {title} {\enquote {\bibinfo {title} {Distinguishing quantum from
  classical {R}abi oscillations in a phase qubit},}\ }\href@noop {} {\bibfield
  {journal} {\bibinfo  {journal} {New J. Phys.}\ }\textbf {\bibinfo {volume}
  {10}},\ \bibinfo {pages} {073026} (\bibinfo {year} {2008})}\BibitemShut
  {NoStop}%
\bibitem [{\citenamefont {Tuorila}\ \emph {et~al.}(2010)\citenamefont
  {Tuorila}, \citenamefont {Silveri}, \citenamefont {Sillanp\"a\"a},
  \citenamefont {Thuneberg}, \citenamefont {Makhlin},\ and\ \citenamefont
  {Hakonen}}]{Tuorila10a}%
  \BibitemOpen
  \bibfield  {author} {\bibinfo {author} {\bibfnamefont {J.}~\bibnamefont
  {Tuorila}}, \bibinfo {author} {\bibfnamefont {M.}~\bibnamefont {Silveri}},
  \bibinfo {author} {\bibfnamefont {M.}~\bibnamefont {Sillanp\"a\"a}}, \bibinfo
  {author} {\bibfnamefont {E.}~\bibnamefont {Thuneberg}}, \bibinfo {author}
  {\bibfnamefont {Y.}~\bibnamefont {Makhlin}}, \ and\ \bibinfo {author}
  {\bibfnamefont {P.}~\bibnamefont {Hakonen}},\ }\bibfield  {title} {\enquote
  {\bibinfo {title} {Stark effect and generalized {Bloch-Siegert} shift in a
  strongly driven two-level system},}\ }\href@noop {} {\bibfield  {journal}
  {\bibinfo  {journal} {Phys. Rev. Lett.}\ }\textbf {\bibinfo {volume} {105}},\
  \bibinfo {pages} {257003} (\bibinfo {year} {2010})}\BibitemShut {NoStop}%
\bibitem [{\citenamefont {Romh\'anyi}\ \emph {et~al.}(2015)\citenamefont
  {Romh\'anyi}, \citenamefont {Burkard},\ and\ \citenamefont
  {P\'alyi}}]{Romhanyi15a}%
  \BibitemOpen
  \bibfield  {author} {\bibinfo {author} {\bibfnamefont {J.}~\bibnamefont
  {Romh\'anyi}}, \bibinfo {author} {\bibfnamefont {G.}~\bibnamefont {Burkard}},
  \ and\ \bibinfo {author} {\bibfnamefont {A.}~\bibnamefont {P\'alyi}},\
  }\bibfield  {title} {\enquote {\bibinfo {title} {Subharmonic transitions and
  {Bloch-Siegert} shift in electrically driven spin resonance},}\ }\href@noop
  {} {\bibfield  {journal} {\bibinfo  {journal} {Phys. Rev. B}\ }\textbf
  {\bibinfo {volume} {92}},\ \bibinfo {pages} {054422} (\bibinfo {year}
  {2015})}\BibitemShut {NoStop}%
\bibitem [{\citenamefont {Shevchenko}\ \emph {et~al.}(2012)\citenamefont
  {Shevchenko}, \citenamefont {Omelyanchouk},\ and\ \citenamefont
  {Il'ichev}}]{Shevchenko12}%
  \BibitemOpen
  \bibfield  {author} {\bibinfo {author} {\bibfnamefont {S.~N.}\ \bibnamefont
  {Shevchenko}}, \bibinfo {author} {\bibfnamefont {A.~N.}\ \bibnamefont
  {Omelyanchouk}}, \ and\ \bibinfo {author} {\bibfnamefont {E.}~\bibnamefont
  {Il'ichev}},\ }\bibfield  {title} {\enquote {\bibinfo {title} {Multiphoton
  transitions in {J}osephson-junction qubits},}\ }\href@noop {} {\bibfield
  {journal} {\bibinfo  {journal} {Low Temp. Phys.}\ }\textbf {\bibinfo {volume}
  {38}},\ \bibinfo {pages} {283} (\bibinfo {year} {2012})}\BibitemShut
  {NoStop}%
\bibitem [{\citenamefont {Grajcar}\ \emph {et~al.}(2008)\citenamefont
  {Grajcar}, \citenamefont {Van~der Ploeg}, \citenamefont {Izmalkov},
  \citenamefont {Il'ichev}, \citenamefont {Meyer}, \citenamefont {Fedorov},
  \citenamefont {Shnirman},\ and\ \citenamefont {Sch{\"o}n}}]{Grajcar08}%
  \BibitemOpen
  \bibfield  {author} {\bibinfo {author} {\bibfnamefont {M.}~\bibnamefont
  {Grajcar}}, \bibinfo {author} {\bibfnamefont {S.~H.~W.}\ \bibnamefont
  {Van~der Ploeg}}, \bibinfo {author} {\bibfnamefont {A.}~\bibnamefont
  {Izmalkov}}, \bibinfo {author} {\bibfnamefont {E.}~\bibnamefont {Il'ichev}},
  \bibinfo {author} {\bibfnamefont {H.-G.}\ \bibnamefont {Meyer}}, \bibinfo
  {author} {\bibfnamefont {A.}~\bibnamefont {Fedorov}}, \bibinfo {author}
  {\bibfnamefont {A.}~\bibnamefont {Shnirman}}, \ and\ \bibinfo {author}
  {\bibfnamefont {G.}~\bibnamefont {Sch{\"o}n}},\ }\bibfield  {title} {\enquote
  {\bibinfo {title} {Sisyphus cooling and amplification by a superconducting
  qubit},}\ }\href@noop {} {\bibfield  {journal} {\bibinfo  {journal} {Nature
  Phys.}\ }\textbf {\bibinfo {volume} {4}},\ \bibinfo {pages} {612} (\bibinfo
  {year} {2008})}\BibitemShut {NoStop}%
\bibitem [{\citenamefont {Nori}(2008)}]{Nori08}%
  \BibitemOpen
  \bibfield  {author} {\bibinfo {author} {\bibfnamefont {F.}~\bibnamefont
  {Nori}},\ }\bibfield  {title} {\enquote {\bibinfo {title} {Superconducting
  qubits: {A}tomic physics with a circuit},}\ }\href@noop {} {\bibfield
  {journal} {\bibinfo  {journal} {Nature Phys.}\ }\textbf {\bibinfo {volume}
  {4}},\ \bibinfo {pages} {589} (\bibinfo {year} {2008})}\BibitemShut {NoStop}%
\bibitem [{\citenamefont {Gonzalez-Zalba}\ \emph {et~al.}(2015)\citenamefont
  {Gonzalez-Zalba}, \citenamefont {Barraud}, \citenamefont {Ferguson},\ and\
  \citenamefont {Betz}}]{Gonzalez15}%
  \BibitemOpen
  \bibfield  {author} {\bibinfo {author} {\bibfnamefont {M.~F.}\ \bibnamefont
  {Gonzalez-Zalba}}, \bibinfo {author} {\bibfnamefont {S.}~\bibnamefont
  {Barraud}}, \bibinfo {author} {\bibfnamefont {A.~J.}\ \bibnamefont
  {Ferguson}}, \ and\ \bibinfo {author} {\bibfnamefont {A.~C.}\ \bibnamefont
  {Betz}},\ }\bibfield  {title} {\enquote {\bibinfo {title} {Probing the limits
  of gate-based charge sensing},}\ }\href@noop {} {\bibfield  {journal}
  {\bibinfo  {journal} {Nature Comm.}\ }\textbf {\bibinfo {volume} {6}},\
  \bibinfo {pages} {6084} (\bibinfo {year} {2015})}\BibitemShut {NoStop}%
\bibitem [{\citenamefont {Shevchenko}\ \emph {et~al.}(2015)\citenamefont
  {Shevchenko}, \citenamefont {Rubanov},\ and\ \citenamefont
  {Nori}}]{Shevchenko15}%
  \BibitemOpen
  \bibfield  {author} {\bibinfo {author} {\bibfnamefont {S.~N.}\ \bibnamefont
  {Shevchenko}}, \bibinfo {author} {\bibfnamefont {D.~G.}\ \bibnamefont
  {Rubanov}}, \ and\ \bibinfo {author} {\bibfnamefont {F.}~\bibnamefont
  {Nori}},\ }\bibfield  {title} {\enquote {\bibinfo {title} {Delayed-response
  quantum back action in nanoelectromechanical systems},}\ }\href@noop {}
  {\bibfield  {journal} {\bibinfo  {journal} {Phys. Rev. B}\ }\textbf {\bibinfo
  {volume} {91}},\ \bibinfo {pages} {165422} (\bibinfo {year}
  {2015})}\BibitemShut {NoStop}%
\bibitem [{\citenamefont {Okazaki}\ \emph {et~al.}(2016)\citenamefont
  {Okazaki}, \citenamefont {Mahboob}, \citenamefont {Onomitsu}, \citenamefont
  {Sasaki},\ and\ \citenamefont {Yamaguchi}}]{Okazaki16}%
  \BibitemOpen
  \bibfield  {author} {\bibinfo {author} {\bibfnamefont {Y.}~\bibnamefont
  {Okazaki}}, \bibinfo {author} {\bibfnamefont {I.}~\bibnamefont {Mahboob}},
  \bibinfo {author} {\bibfnamefont {K.}~\bibnamefont {Onomitsu}}, \bibinfo
  {author} {\bibfnamefont {S.}~\bibnamefont {Sasaki}}, \ and\ \bibinfo {author}
  {\bibfnamefont {H.}~\bibnamefont {Yamaguchi}},\ }\bibfield  {title} {\enquote
  {\bibinfo {title} {Gate-controlled electromechanical backaction induced by a
  quantum dot},}\ }\href@noop {} {\bibfield  {journal} {\bibinfo  {journal}
  {Nat. Commun.}\ }\textbf {\bibinfo {volume} {7}},\ \bibinfo {pages} {11132}
  (\bibinfo {year} {2016})}\BibitemShut {NoStop}%
\bibitem [{\citenamefont {Josephson}(1962)}]{Josephson62}%
  \BibitemOpen
  \bibfield  {author} {\bibinfo {author} {\bibfnamefont {B.~D.}\ \bibnamefont
  {Josephson}},\ }\bibfield  {title} {\enquote {\bibinfo {title} {Possible new
  effects in superconductive tunnelling},}\ }\href@noop {} {\bibfield
  {journal} {\bibinfo  {journal} {Phys. Lett.}\ }\textbf {\bibinfo {volume}
  {1}},\ \bibinfo {pages} {251} (\bibinfo {year} {1962})}\BibitemShut {NoStop}%
\bibitem [{\citenamefont {Zhou}\ \emph {et~al.}(2014)\citenamefont {Zhou},
  \citenamefont {Huang}, \citenamefont {Zhang}, \citenamefont {Wang},
  \citenamefont {Tan}, \citenamefont {Xu}, \citenamefont {Shi}, \citenamefont
  {Rong}, \citenamefont {Ashhab},\ and\ \citenamefont {Du}}]{Zhou14}%
  \BibitemOpen
  \bibfield  {author} {\bibinfo {author} {\bibfnamefont {J.}~\bibnamefont
  {Zhou}}, \bibinfo {author} {\bibfnamefont {P.}~\bibnamefont {Huang}},
  \bibinfo {author} {\bibfnamefont {Q.}~\bibnamefont {Zhang}}, \bibinfo
  {author} {\bibfnamefont {Z.}~\bibnamefont {Wang}}, \bibinfo {author}
  {\bibfnamefont {T.}~\bibnamefont {Tan}}, \bibinfo {author} {\bibfnamefont
  {X.}~\bibnamefont {Xu}}, \bibinfo {author} {\bibfnamefont {F.}~\bibnamefont
  {Shi}}, \bibinfo {author} {\bibfnamefont {X.}~\bibnamefont {Rong}}, \bibinfo
  {author} {\bibfnamefont {S.}~\bibnamefont {Ashhab}}, \ and\ \bibinfo {author}
  {\bibfnamefont {J.}~\bibnamefont {Du}},\ }\bibfield  {title} {\enquote
  {\bibinfo {title} {Observation of time-domain {R}abi oscillations in the
  {Landau-Zener} regime with a single electronic spin},}\ }\href@noop {}
  {\bibfield  {journal} {\bibinfo  {journal} {Phys. Rev. Lett.}\ }\textbf
  {\bibinfo {volume} {112}},\ \bibinfo {pages} {010503} (\bibinfo {year}
  {2014})}\BibitemShut {NoStop}%
\bibitem [{\citenamefont {Landau}\ and\ \citenamefont
  {Lifshitz}(1981)}]{Landau1981QM}%
  \BibitemOpen
  \bibfield  {author} {\bibinfo {author} {\bibfnamefont {L.~D.}\ \bibnamefont
  {Landau}}\ and\ \bibinfo {author} {\bibfnamefont {L.~M.}\ \bibnamefont
  {Lifshitz}},\ }\href@noop {} {\emph {\bibinfo {title} {{Course of Theoretical
  Physics, Volume III: Quantum Mechanics (Non-Relativistic Theory)}}}},\
  \bibinfo {edition} {3rd}\ ed.\ (\bibinfo  {publisher}
  {Butterworth-Heinemann},\ \bibinfo {year} {1981})\BibitemShut {NoStop}%
\bibitem [{\citenamefont {Pershin}\ \emph {et~al.}(2015)\citenamefont
  {Pershin}, \citenamefont {Shevchenko},\ and\ \citenamefont
  {Nori}}]{Pershin15}%
  \BibitemOpen
  \bibfield  {author} {\bibinfo {author} {\bibfnamefont {Y.~V.}\ \bibnamefont
  {Pershin}}, \bibinfo {author} {\bibfnamefont {S.~N.}\ \bibnamefont
  {Shevchenko}}, \ and\ \bibinfo {author} {\bibfnamefont {F.}~\bibnamefont
  {Nori}},\ }\bibfield  {title} {\enquote {\bibinfo {title} {Memcapacitors and
  meminductors based on superconducting qubits},}\ }in\ \href@noop {} {\emph
  {\bibinfo {booktitle} {Proceedings of the International Symposium on
  Nanoscale Transport and Technology (ISNTT2015), NTT Atsugi R\&D Center,
  Atsugi, Japan, November 17-20}}}\ (\bibinfo {year} {2015})\ p.~\bibinfo
  {pages} {72}\BibitemShut {NoStop}%
\end{thebibliography}%

\end{document}